\documentclass[a4paper,noarxiv,twocolumn]{quantumarticle}

\usepackage[utf8]{inputenc}
\usepackage[english]{babel}
\usepackage[T1]{fontenc}
\usepackage[colorlinks=true,linkcolor=quantumviolet,citecolor=quantumviolet,urlcolor=quantumviolet]{hyperref}

\usepackage[numbers,sort&compress]{natbib}

\usepackage{ltxgrid}

\usepackage{tikz}
\usepackage{lipsum}

\usepackage{dcolumn}
\usepackage{bm}
\usepackage{physics}
\usepackage[ruled,lined]{algorithm2e}
\usepackage{algpseudocode}
\usepackage{algpseudocode}
\usepackage{amsmath}
\usepackage{amsfonts}
\usepackage{amssymb,dsfont,amsthm}
\usepackage{subfigure}
\usepackage{subfigure}
\usepackage{float}

\newtheorem{thm}{Theorem}

\theoremstyle{definition}

\newcommand\identity{1\kern-0.25em\text{l}}
\newcommand{\Rh}{\hat{\bm{\rho}}}
\newcommand{\Povm}{\hat{\Pi}}

\newcommand{\complex}{\mathbb{C}}

\newcommand{\diag}{\mathrm{diag}}
\makeatletter
\newcommand{\ALOOP}[1]{\ALC@it\algorithmicloop\ #1%
  \begin{ALC@loop}}
\newcommand{\ENDALOOP}{\end{ALC@loop}\ALC@it\algorithmicendloop}

\makeatother

\begin{document}

\title{Resourcefulness of non-classical continuous-variable quantum gates}

\author{Massimo Frigerio}
\affiliation{Laboratoire Kastler Brossel, Sorbonne Universit\'{e}, CNRS, ENS-Universit\'{e} PSL,  Coll\`{e}ge de France, 4 place Jussieu, F-75252 Paris, France}
\author{Antoine Debray}
\affiliation{Laboratoire Kastler Brossel, Sorbonne Universit\'{e}, CNRS, ENS-Universit\'{e} PSL,  Coll\`{e}ge de France, 4 place Jussieu, F-75252 Paris, France}
\author{Nicolas Treps}
\affiliation{Laboratoire Kastler Brossel, Sorbonne Universit\'{e}, CNRS, ENS-Universit\'{e} PSL,  Coll\`{e}ge de France, 4 place Jussieu, F-75252 Paris, France}
\author{Mattia Walschaers}
\affiliation{Laboratoire Kastler Brossel, Sorbonne Universit\'{e}, CNRS, ENS-Universit\'{e} PSL,  Coll\`{e}ge de France, 4 place Jussieu, F-75252 Paris, France}
\email{mattia.walschaers@lkb.upmc.fr}
\date{\today}

\begin{abstract}
In continuous-variable quantum computation, identifying key elements that enable a quantum computational advantage is a long-standing issue. Starting from the standard results on the necessity of Wigner negativity, we develop a comprehensive and versatile approach in which the techniques of $(s)$-ordered quasiprobabilities are exploited to provide rigorous statements on the simulability of photonic quantum circuits consisting of previously characterized gates and thereby identifying the contribution of each quantum gate to the potential achievement of quantum computational advantage. This is achieved by means of an analysis of the so-called transfer function, allowing us to highlight the resourcefulness of a gate set. As such this technique can be straightforwardly applied to current continuous-variables quantum circuits, while also constraining the tolerable amount of losses above which any potential quantum advantage can be ruled out. We use $(s)$-ordered quasiprobability distributions on phase-space to capture the non-classical features in the protocol, and focus our technique entirely on the ordering parameter $s$. This allows us to highlight the resourcefulness and robustness to loss of a universal set of unitary gates comprising three distinct Gaussian gates and any non-Gaussian unitary gate, providing important insight on the role of non-Gaussianity.
\end{abstract}

\maketitle

\section{Introduction}
Ever since Shor's algorithm \cite{doi:10.1137/S0036144598347011} promised an exponential speed-up in the factorisation of prime numbers, quantum computing has captivated the scientific community. The laws and logic behind quantum physics make it possible to solve certain highly specific problems much more efficiently than what could be achieved with classical devices. While it is not unreasonable to expect that quantum computers hold huge potential as a new computational framework, so far actual algorithmic design has only scratched the surface \cite{PRXQuantum.2.040203}.

A bit over a decade ago, a new approach to study the question of the computational advantage of quantum devices over classical ones was introduced. Specific protocols such as boson sampling \cite{10.1145/1993636.1993682}, IQP \cite{doi:10.1098/rspa.2010.0301}, and random circuit sampling \cite{RCS}, were proposed as candidates for reaching a provable quantum computational advantage. The foundations of these works within computational complexity theory \cite{harrow2017quantum,RevModPhys.95.035001} separates them from earlier works on quantum speed-ups in the sense that there are strong arguments to suggest that these protocols cannot be simulated efficiently by any classical algorithm. These works have ultimately led to a series of experiments claiming to have reached such a quantum computational advantage \cite{QuantumSupremacy,doi:10.1126/science.abe8770,madsen2022quantum, decross2024computationalpowerrandomquantum}. 

In the field of photonic quantum computation, the boson sampling protocol and its many variations \cite{PhysRevLett.119.170501, PhysRevLett.113.100502, PhysRevA.96.062307, PhysRevA.96.032326,quesada2018gaussian} have drawn much attention. The race for a quantum computational advantage quickly imposed the question: when do noise and experimental imperfections make a setup efficiently simulable on a classical computer? This has led to applied results exploiting the fine details of boson sampling \cite{garcia2019simulating,renema2020simulability,PhysRevLett.120.220502, rahimi2016sufficient} and Gaussian boson sampling \cite{doi:10.1126/sciadv.abl9236,PhysRevLett.124.100502,PRXQuantum.3.010306,PhysRevA.105.012427,MPS-sim} setups. Furthermore, this sparked research activities on the simulation of photonic sampling problems in a more general context. 

This more general line of research can be seen as a series of attempts to translate and generalise the Gottesman-Knill theorem \cite{gottesman1998heisenbergrepresentationquantumcomputers} for bosonic systems. The first such generalisation dates back to the early days of continuous-variable quantum information processing, and it showed simply that a setup involving exclusively Gaussian states, operations, and measurements can be efficiently simulated \cite{Bartlett2002}. For a highly ideal setup with pure states, unitary gates, and projective measurements, Hudson's theorem \cite{hudson1974wigner,soto1983wigner} teaches us that these Gaussian elements are the only ones that are described by a fully positive Wigner function \cite{cahill1969density}. When more general sampling setups are considered, allowing states to be mixed, measurements can be general positive operator-valued measure (POVMs), and gates to be described by general channels, it turns out that this positivity of the Wigner function is the element that makes a system efficiently simulable \cite{mari2012positive,veitch2012negative}. In other words, the existence of negative regions in the Wigner functions that describe the system is a key resource for reaching a quantum computational advantage.

While such Wigner negativity might be a necessary resource, it clearly is not a sufficient one \cite{garcia2020efficient}. A series of recent works has used tools from bosonic codes to completely challenge the paradigm of Wigner negativity as a crucial resource \cite{Calcluth2022efficientsimulation, PhysRevA.107.062414}, relying instead on discrete-variable resources that come from the qubits that bosonic codes are encoding \cite{PRXQuantum.5.020337}. However, these sufficient resources may turn out to be more demanding than necessary. In this regard, other works have searched for new necessary resources for reaching a quantum computational advantage, leading to the identification of stellar rank and some non-Gaussian type of entanglement as key ingredients \cite{PhysRevLett.130.090602}. Other approaches have aimed to generalise the results of Wigner negativity to more general phase-space representations \cite{rahimi2016sufficient, chabaud2024phasespacenegativity}.

Most of these recent results have focused primarily on the resources that are present in the initial states and/or the measurements. The resourcefulness of the operations that are implemented in between (typically quantum gates) have not received much attention. In works such as \cite{Calcluth2022efficientsimulation, rahimi2016sufficient} these operations are limited to a very specific class of circuits, or one directly considers measurements on some generic state as in \cite{PhysRevLett.130.090602}. These works therefore assume, either explicitly or implicitly, that a straightforward approach can be taken when addressing big multimode channels connecting input states to measurements. This is in strong contrast to \cite{mari2012positive}, where a lot of attention is devoted to the development of a Markov-chain model to simulate the action of a sequence of few-mode quantum channels. This can effectively be understood as a hidden-variable model for quantum computing which allows, at least in principle, to study a very wide set of operations.

In the present work, we provide insight and results on the simulatability of quantum gates by arguing, in contrast to \cite{rahimi2016sufficient}, that the ordering parameters can be chosen in an optimal way that also allows to treat quantum gates in a decoupled, independent manner. Our gate-by-gate model allows for the precise identification of the sources of quantum computational advantage within a circuit and its tolerance to local losses. As such, we introduce a algorithmic approach to derive the best decomposition of a quantum circuit in terms of $(s)$-ordered quasiprobability distributions, we demonstrate the significant insights that this can yield. It is worth highlighting that our approach effectively looks for an optimal hidden-variable model for the computation, by optimizing the choice of $s$-ordering parameters in the representation of subsequent layers of operations. Moreover, this algorithm allows us to effectively deal with losses or noise at precise points in the circuit. Even if we find that a certain protocol is impossible to simulate efficiently, because no appropriate $(s)$-parameters can be found, our approach allows us to identify how much loss or noise can be tolerated before the setup can be efficiently simulated. 

The key results of our work lie in the second part, Sec. \ref{sec:Gates} and Sec. \ref{sec5}, where we devote attention to specific types of quantum gates and channels that are highly relevant in photonic quantum computing. Through an examination of their respective transfer functions, and the manner in which each function fits within the decoupled approach outlined in section \ref{sec3}, it becomes possible to pinpoint the precise sources of quantum computational advantage within a circuit, or, conversely, where such advantage is crucially lost. This notably allows to understand the role of squeezing operations, photon addition and subtraction, and any deterministic non-Gaussian gate, such as the cubic phase gate, in reaching a quantum computational advantage.

We stress that the primary aim of this work is not to introduce a new classical simulation algorithm, but rather to establish rigorous, gate-level conditions under which a continuous-variable quantum computation can be efficiently simulated classically. Algorithm 1 serves as a concise formalism to state and organise these conditions: by expressing the simulability of the full circuit as the conjunction of local positivity constraints on (s)-ordered transfer functions, it reduces a global problem to a tractable sequence of single-gate analyses. It is in this analytical reduction, and in the concrete characterisation of the resulting conditions for a universal gate set, that the contribution of this work lies.

\section{Phase-space Representations}
\subsection{$(s)$-ordered representations of states and observables}
\begin{figure}
    \centering
    \includegraphics[scale=0.45]{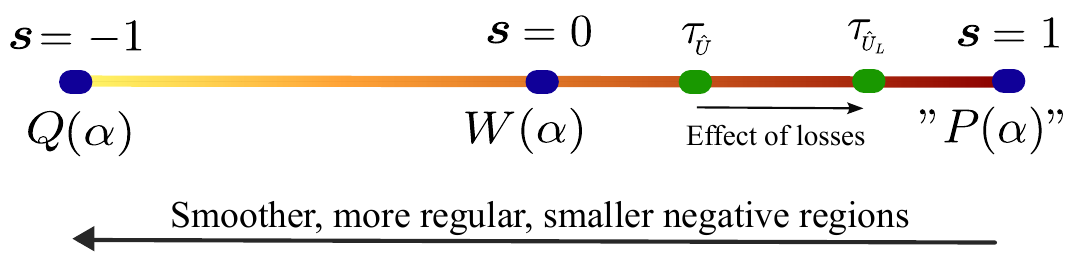}
    \caption{The range of values of the $(s)$-parameter. The values corresponding to the $Q-$, $P-$, and Wigner function are highlighted. We also visually represent the effect of Eq.~\eqref{losses_result}, which describes how losses drive the nonclassical parameter closer to one. }
    \label{fig:losses-one}
\end{figure}

The main theoretical tool on which our results rely consists of $(s)$-ordered quasiprobability distributions (quasi-PDF) on phase-space. 
Let us consider a Fock space of $M$ bosonic modes, each associated with its pair of creation and annihilation operators $\hat{a}_{j}^{\dagger}, \hat{a}_j$ such that $[ \hat{a}_j , \hat{a}_{k}^\dagger ] = \delta_{jk}$. We denote by $\hat{\bm{a}} = (\hat{a}_{1}, ..., \hat{a}_M)$ the corresponding vector of operators.
For a bounded operator $\hat O$ on such space, and for any real number $s\in [-1,1]$, we then define the $(s)$-ordered phase-space representation \cite{cahill1969density, cahill1969ordered}
\begin{equation}\label{eq:sParameterDef}
    W_{\hat O}^{(s)}(\bm{\alpha}) = \int_{\bm{\xi} \in \mathbb{C}^M}\frac{d^{2M}\bm{\xi}}{\pi^{2M}} \chi_{\hat O}^{(s)}(\bm{\xi})e^{\bm{\alpha} \bm{\xi}^\dag - \bm{\xi} \bm{\alpha}^\dag},
\end{equation}
where $\bm{\xi} , \bm{\alpha} \in \complex^M$ are complex vectors. The $(s)$-ordered characteristic function $\chi_{\hat O}^{(s)}(\bm{\xi})$ is defined by 
\begin{equation}\label{eq:CharacteristicFunction}
    \chi_{\hat O}^{(s)}(\xi) = \Tr[\hat O \hat D_{\bm{s}} (\bm{\xi})],
\end{equation}
with $\hat D_{\bm{s}} (\bm{\xi})$ the $(\bm{s})$-ordered displacement operator $\hat{D}_{\bm{s}}(\bm{\xi})= \hat{D}(\bm{\xi})e^{\bm{\xi}\bm{s} \bm{\xi}^\dag/2}$, which is a generalisation of the standard $M$-modes displacement operator $ \hat{D}(\bm{\xi}) = \exp { \bm{\xi} \hat{\bm{a}}^\dagger - \bm{\xi}^\dagger \hat{\bm{a}} }.$ These $(s)$-ordered representations are useful tools to study CV quantum systems due to the identity \cite{cahill1969density, cahill1969ordered}
    \begin{equation}
    \label{eq:swithminuss}
        \Tr[\hat O_1 \hat O_2] = \int_{\bm{\alpha} \in \mathbb{C}^M} \frac{d^{2M}\bm{\alpha}}{\pi^M} W_{\hat O_1}^{(-s)}(\bm{\alpha})W_{\hat O_2}^{(s)}(\bm{\alpha}).
    \end{equation}
We emphasize the requirement of pairing a quasiprobabiliy distribution with parameter $s$ with one with parameter $-s$. As indicated in Fig.~\ref{fig:losses-one}, the higher the value of $s$, the more singular the distributions. Vice-versa, lower values of $s$ indicate more regular phase-space representations. 

Phase-space representations are often used to characterise quantum states, in which case we choose $\hat O$ in Eq.~\eqref{eq:sParameterDef} to be the density matrix $\hat \rho$. The $(s)$-ordered phase-space representation then becomes a quasiprobability distribution, meaning that it is normalised, but not necessarily positive. For the case $s=0$, we recover the Wigner function. For $s=-1$, instead, we have the Husimi Q-function, which is always non-negative and normalised to $1$, being a proper probability density function (PDF) associated with non-orthogonal projectors onto coherent states. On the opposite hand, for $s=1$, the integral won't always converge to some regular function and the result is instead defined as a distribution, i.e. a linear functional on a space of test functions, and it bears the name of Glauber's $P$-representation. 

Importantly, it is possible to transform a $(s)$-ordered representation into a $(s')$-ordered representation with $s' < s$ by convoluting it with a Gaussian function:
\begin{equation}
\label{eq:Wttot}
                W^{(s')}_{\hat{O}} ( \alpha ) \ = \ \dfrac{2}{s-s'} \int_{ \beta \in \mathbb{C}} \frac{d^{2} \beta}{\pi} e^{ -\frac{ 2 \vert \alpha - \beta \vert^2}{s - s'}} W^{(s)}_{\hat{O}} ( \beta ).
\end{equation}
Clearly then, $W^{(s')}_{\hat{O}} (\alpha)$ will be more regular than $ W^{(s)}_{\hat{O}} ( \alpha )$ for $s' < s$ \cite{PhysRevA.51.3340}. If we are dealing with a quantum state, $\hat{O} = \Rh$, with a negative Wigner function ($s=0$), looking for the highest value of $s' < 0$ such that $W^{(s')}_{\Rh}$ is a PDF amounts to finding the narrowest Gaussian kernel that will erase all the negative regions in the original Wigner function. As such, it is a means of quantifying Wigner negativity, referred to as the non-classical depth, which will be addressed in the following paragraph. Unlike most common measures of Wigner negativity, such as the negativity volume, though, it also takes into account how the negative regions are shaped in phase-space: a very elongated negativity region would require a much broader Gaussian to be fully cancelled, when compared to another region with the same negative volume but concentrated in a more symmetric area. To quantify this idea, we can use the negativity volume:
\begin{equation}
\mathcal{N}^{(s)}(\hat \rho) = \int_{\bm{\alpha} \in \mathbb{C}^M} d^{2M}\bm{\alpha} \lvert W_{\hat \rho}^{(s)}(\bm{\alpha})\rvert - 1.
\end{equation}
Combining with Eq.~\eqref{eq:Wttot} and using the fact that the modulus of an integral is always majorized by the integral of the modulus, we directly find that
\begin{equation}\label{eq:negativityIneq}
\mathcal{N}^{(s')}(\hat \rho) \leq \mathcal{N}^{(s)}(\hat \rho) \quad \text{for all $s' \leq s$}.
\end{equation}
More generally, also for Wigner-positive states, we can define $\tau$ as the maximal value for which $\mathcal{N}^{(\tau)}(\hat \rho) = 0,$ and thus $ W^{(\tau)}_{\hat{\rho}} ( \alpha )$ is a PDF. Given that the Husimi Q-function is obtained by projecting $\hat \rho$ onto coherent states, it is clear that such a $\tau \geq -1$ always exists. This value $\tau$ defines the \emph{nonclassical depth} \cite{lee1991measure} of the state $\hat \rho$:
\begin{equation}
\mathfrak{t}  [ \Rh ] \ \ := \ \ \frac{ 1 - \tau}{2}
\end{equation}
All non-Gaussian pure states have nonclassical depth $\mathfrak{t} = 1$ (meaning that only their $Q$-function is a PDF). For Gaussian states, since they have positive Wigner functions, $\mathfrak{t} \leq 1/2$, and for the particular case of coherent states we obtain $\mathfrak{t} = 0$. The latter are the only pure states whose $P$-function is a PDF, thus they are commonly referred to as the only classical pure state. The nonclassical depth can be employed as a measure of nonclassicality of quantum states and is particularly informative for mixed states and noisy quantum processes.

Up to now, we have only considered the case with a single ordering parameter. However, dealing with multimode states, it is much fruitful to define the $(\bm{s})$-ordered phase-space representation such that $\bm{s} = \mathrm{diag}(s_1, s_2, \dots, s_M)$ is a diagonal matrix, in such a way that each mode has its own independent ordering parameter, instead of choosing a single ordering parameter for all modes. This simply boils down to replacing $\hat{D}_{\bm{s}} (\bm{\xi})$ with $\hat{D}_{\bm{s}} (\bm{\xi})= \hat{D}(\bm{\xi})e^{\bm{\xi}\bm{s}\bm{\xi}^\dag/2}$ in Eq.~\eqref{eq:CharacteristicFunction}. Consequently, the $(\bm{s})$-ordered representation of any operator $\hat O$ is designated by $W^{(\bm{s})}_{\hat O}$. For $\bm{s}=-\identity_M$, $\bm{s}=0_M$ and $\bm{s}= \identity_M$ one finds respectively the Husimi $Q$-function, the Wigner function and the Glauber-Sudarshan $P$-function. 

\subsection{$(s)$-ordered representations of quantum operations}

We first consider a generic quantum optical scheme:  $M$ input bosonic modes, described by a density operator $\Rh_{in}$, are processed by a trace-preserving completely positive (CPTP) quantum channel $\mathcal{E}$. The overall output state $\Rh_{out} = \mathcal{E}(\Rh_{in})$ is then measured by a \emph{factorized} POVM $\Povm_{\bm{x}}$, $\bm{x}$ being the generic label for each possible measurement result. It was shown in \cite{rahimi2016sufficient} that the probability to obtain the outcome $\bm{x}$ is given by
\begin{multline}\label{eq:probaCaves}
        p(\bm{x}) = \iint_{\bm{\alpha}, \bm{\beta} \in \mathbb{C}^M} \pi^{M}d^{2M}\bm{\alpha}d^{2M}\bm{\beta}\  W_{\Rh_{\mathrm{in}}}^{(\bm{s'})}(\bm{\alpha}) \\ \times \Lambda_{\mathcal{E}}^{(-\bm{s'},\bm{s})}(\bm{\alpha},\bm{\beta}) W_{\Povm_{\bm{x}}}^{(-\bm{s})}(\bm{\beta}),
\end{multline}
where $\Lambda_{\mathcal{E}}^{(\bm{s'},\bm{s})}(\alpha,\beta)$ is the $(\bm{s'},\bm{s})$-ordered transfer function of $\mathcal{E}$, defined as: 
\begin{multline}\label{eq:transferFunction}
    \Lambda_{\mathcal{E}}^{(\bm{s'},\bm{s})}(\bm{\alpha},\bm{\beta}) = \\ \iint_{\bm{\zeta}, \bm{\xi}  \in \mathbb{C}^M} \frac{d^{2M}\bm{\zeta}}{\pi^M} \frac{d^{2M}\bm{\xi}}{\pi^{2M}} \ e^{\bm{\beta} \bm{\zeta}^\dag - \bm{\zeta}\bm{\beta}^\dag+\bm{\xi}\bm{\alpha}^\dag - \bm{\alpha} \bm{\xi}^\dag}  \\ \,\times e^{\bm{\zeta}\bm{s}\bm{\zeta}^\dag/2 + \bm{\xi}\bm{s'}\bm{\xi}^\dag/2} \Tr[\mathcal{E}(\hat{D}^\dag(\bm{\xi}))\hat{D} (\bm{\zeta})] .
\end{multline}
As such, the transfer function $\Lambda_{\mathcal{E}}^{(\bm{s'},\bm{s})}(\alpha,\beta)$ is a phase-space representation of the trace-preserving quantum operation $\mathcal{E}$. Moreover, the transfer function in Eq.~\eqref{eq:transferFunction} comes with an interesting additional property that derives from the fact that it represents a CPTP map:
\begin{equation}\label{eq:CPTPCondition}
\int_{ \bm{\beta} \in \mathbb{C}^M} d^{2M}\bm{\beta} \,\Lambda_{\mathcal{E}}^{(\bm{s'},\bm{s})}(\bm{\alpha},\bm{\beta}) = 1.
\end{equation}
In Section \ref{sec:algorithm}, we will use that \eqref{eq:CPTPCondition} implies, whenever $\Lambda_{\mathcal{E}_i}^{(\bm{s'}, \bm{s})}$ is a regular and positive function, that we can interpret $\Lambda_{\mathcal{E}_i}^{(\bm{s'}, \bm{s})}(\bm{\alpha},\bm{\beta})$ as a probability distribution for $\bm{\beta}$, for every possible value of $\bm{\alpha}$.

For a general multimode system Eq.~\eqref{eq:transferFunction} is usually very hard to work out explicitly. This should not come as a surprise, given that, for setups with factorized multi-mode input states and single-mode measurements, the map $\mathcal{E}$ should provide the necessary entanglement to make the setup hard to simulate on a classical computer. It is therefore common to make some assumptions on $\mathcal{E}$.\\

In \cite{rahimi2016sufficient}, the authors focus on boson sampling and, thus, concentrate primarily on the case where $\mathcal{E}_{L}$ is a passive linear-optics channel. In this case, it is possible to find an analytical expression for $\Lambda_{\mathcal{E}_L}^{(\bm{s'},\bm{s})}(\bm{\alpha},\bm{\beta})$ by using that $\mathcal{E}_L(\hat{D}^\dag(\bm{\xi})) = \hat{D}^\dag(L^{\dag} \bm{\xi})$. Here $L$ is an $M \times M$ matrix with $L^{\dag}L \leq \identity_M$ that captures, both, the mixing of the input modes and losses. This led to the key result that $\Lambda_{\mathcal{E}_L}^{(\bm{s'},\bm{s})}(\bm{\alpha},\bm{\beta})$ is positive if and only if $\bm{s}$ and $\bm{s'}$ satisfy the following condition
\begin{equation}
    \label{conditionCaves}
    \identity_M - L^\dag L - \bm{s} + L^\dag \bm{s'} L \geq 0.
\end{equation}
When it is possible to find $\bm{s}$ and $\bm{s'}$ such that condition \eqref{conditionCaves} is satisfied, and such that $W_{\Rh_{\mathrm{in}}}^{(\bm{s'})}(\bm{\alpha})$ and $W_{\Povm_{\bm{x}}}^{(-\bm{s})}(\bm{\beta})$ are both positive and can therefore be treated as PDFs, one can efficiently simulate the sampling process.

Despite the interest of this result, the class of operations that can be described by such an $\mathcal{E}_{L}$ is rather limited by the fact that it is a linear-optical operation; in contrast, when we aim at studying quantum advantage in CV systems, it is crucial to include a wider range of operations that can induce squeezing and non-Gaussianity.\\

In this work, we will also make an assumption on $\mathcal{E}$, namely that the total channel is decomposed into many operations: $\mathcal{E} = \mathcal{E}_1 \circ \mathcal{E}_2 \circ \dots \circ \mathcal{E}_k$. In the context of quantum computing, we typically understand this as a decomposition into quantum gates, i.e. other CPTP quantum channels involving at most $n\ll M$ modes each.  In Section \ref{sec:Gates} we restrict to the common cases of single- and two-mode operations, but the results of Section \ref{sec:algorithm} can be applied more generally. To study this sequence of operations, we rely on the composition rule
\begin{multline}
        \Lambda_{\mathcal{E}_1 \circ \mathcal{E}_2 }^{(\bm{s'},\bm{s})}(\bm{\alpha},\bm{\beta})  \ = \\
        \int_{\bm{\gamma} \in \mathbb{C}^M}d^{2M}\bm{\gamma} \Lambda_{\mathcal{E}_1}^{(\bm{s'},\bm{s''})}(\bm{\alpha},\bm{\gamma})\Lambda_{\mathcal{E}_2 }^{(-\bm{s''},\bm{s})}(\bm{\gamma},\bm{\beta}).
\end{multline}
The fact that an operation only acts on $n$ modes is reflected by the fact that these functions involve at most $n$ components of $\bm{\alpha}, \bm{\beta},$ and $\bm{\gamma}$ each, and behave like delta distributions on the remaining $M-n$ modes. Notice that $s''$ drops out from the left-hand side of the previous equation, according to the cancellation of the terms $e^{ \pm s'' \vert \gamma \vert^2 /2}$ in the $(\pm s'')$-ordered displacement operators appearing in Eq. \eqref{eq:transferFunction}.

\section{Simulation of Sampling Algorithms}\label{sec:algorithm}
\label{sec3}

\begin{figure}
    \centering
    \includegraphics[scale=0.4]{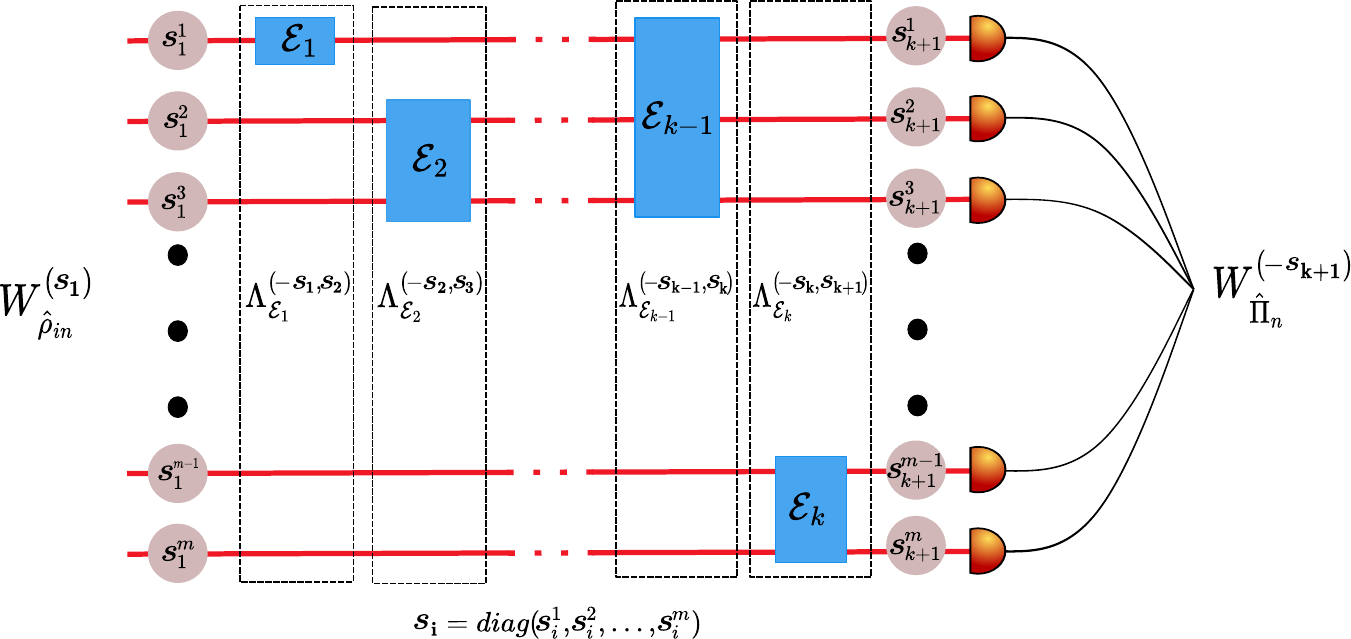}
    \caption{Generic quantum-optical scheme depicted by $M$ input modes, described by a density operator $\Rh_{\mathrm{in}}$ processed through a trace-preserving quantum channel $\mathcal{E}$, that can  be decomposed into a sequence of trace-preserving quantum channels $\mathcal{E} = \mathcal{E}_1 \circ \mathcal{E}_2 \circ \dots \circ \mathcal{E}_k$. This produces the output state $\Rh_{out} = \mathcal{E}(\Rh_{in})$ and an output probability distribution $p(\bm{x}) = \Tr [\Rh_{out}\Povm_{\bm{x}}]$ sampled by measuring the POVM $\Povm_{\bm{x}}$.}
    \label{fig:inter}
\end{figure}

In Fig.~\ref{fig:inter}, we show a schematic representation of the full sampling protocol that is considered in our work: a factorised input state $\Rh_{in} = \hat \rho_1 \otimes \dots \otimes \hat \rho_M$ is sent through a CPTP channel $\mathcal{E} = \mathcal{E}_1 \circ \mathcal{E}_2 \circ \dots \circ \mathcal{E}_k$, and is ultimately measured by a set of single-mode detectors with POVM elements $\Povm_{\bm{x}} = \Povm_{x_1} \otimes \dots \otimes \Povm_{x_M}$. In phase-space, we can generalise Eq. (\ref{eq:probaCaves}) to describe the probability to obtain a string of measurement outcomes $\bm x = (x_1, \dots x_M)$, associated with POVM element $\Povm_{\bm{x}}$, as
\begin{widetext}
\begin{equation}
\begin{aligned}
\label{proba_distrib_sequence}
    p(\bm{x}) \ \ = \ \ \pi^M \int \dots  \int W_{\bm{\hat \rho}_{in}}^{(\bm{s}_1)}(\bm{\alpha}_1) \ &\Lambda_{\mathcal{E}_1}^{(-\bm{s}_1, \bm{s}_2)}(\bm{\alpha}_1,\bm{\alpha}_2) \ \Lambda_{\mathcal{E}_2}^{(-\bm{s}_2, \bm{s}_3)}(\bm{\alpha}_2,\bm{\alpha}_3) \ \dots \\
     \dots \ & \Lambda_{\mathcal{E}_k}^{(-\bm{s}_k, \bm{s}_{k+1})}(\bm{\alpha}_k,\bm{\alpha}_{k+1}) \ W_{\Povm_{\bm{x}}}^{(-\bm{s}_{k+1})}(\bm{\alpha}_{k+1}) \ d\bm{\alpha}_1 \dots d\bm{\alpha}_{k+1}.
    \end{aligned}
\end{equation}
\end{widetext}
Explicit calculation for Eq. \eqref{proba_distrib_sequence} can be found in Appendix (\ref{app:A}). Here, the factorized input state $\Rh_{in}$ of $M$ modes can be represented by a factorized $(\bm{s}_1)$-ordered quasi-PDF $W_{\bm{\hat\rho}_{in}}^{(\bm{s}_1)}(\bm{\alpha}_1)$, and the measurement process is described by a quasi-PDF $W_{\hat{\Pi}_x}^{(-\bm{s}_{k+1})}(\bm{\alpha}_{k+1})$.

If there exists a set $\{\bm{s}_i\}_{i\in[1,k+1]}$ so that each phase-space representation appearing in the decomposition \eqref{proba_distrib_sequence} is regular and positive, the whole scheme can be classically simulated in an efficient way, using the same strategy as the one followed in \cite{mari2012positive}, based on the fact that we can perform each step by sampling a probability distribution (see the second part of Algorithm \ref{alg:main}). We thus seek to optimise each $\bm{s}_i$, which will eventually lead us to Algorithm \ref{alg:main} to check whether a scheme that follows the general structure of Fig.~\ref{fig:inter} can be efficiently simulated by classical means.

We note that the gate-by-gate decomposition provides sufficient conditions for classical simulability: if positivity is achieved at every step, the full circuit is simulable. Conversely, the failure of Algorithm 1 does not imply that the circuit is non-simulable, since a different decomposition or a global optimization over all ordering parameters might still succeed. The trade-off is between tractability — our local conditions are analytically or efficiently computable for each characterised gate — and optimality, which would require a global search over an exponentially large parameter space. As discussed in the outlook, regrouping gates whose individual transfer functions are negative into larger blocks is a natural strategy to tighten the conditions, at the cost of increasing the dimensionality of the sampling space. One should keep into account that simulating the full unitary is believed to be hard task under the assumption that quantum computation is hard to simulate.

To explain this optimisation of the $\bm{s}_i$ parameters, we start with the input state. By virtue of Eq. (\eqref{eq:negativityIneq}), for any single-mode state $\Rh$ there exists a maximal value $\tau_{\Rh} \in [-1,+1]$ such that $W_{\Rh}^{(s)}$ is a probability-density function for any $s \leq \tau_{\Rh}$. The positivity of the Husimi Q-representation on top guarantees that such a parameter $\tau_{\Rh}$ can always be found.

The study of the transfer functions $\Lambda_{\mathcal{E}_i}^{(-\bm{s}_i, \bm{s}_{i+1})}(\bm{\alpha}_i,\bm{\alpha}_{i+1})$ is somewhat more subtle. We note, first of all, that we can easily derive a result similar to Eq.~\eqref{eq:Wttot} for $\bm{s'}_i \succ \bm{s}_i $ and $\bm{s'}_{i+1} \prec \bm{s}_{i+1}$ (as matrix inequalities):
\begin{multline}
    \label{eq:Conv}
\Lambda_{\mathcal{E}_i}^{(-\bm{s'}_i, \bm{s'}_{i+1})}(\bm{\alpha}_i,\bm{\alpha}_{i+1}) = \dfrac{2}{\det(\bm{s}_{i+1}-\bm{s'}_{i+1})\det(\bm{s}'_{i}-\bm{s}_{i})}\\ \times\iint_{ \bm{\beta},\bm{\gamma} \in \mathbb{C}^M} \frac{d^{2M} \bm\beta d^{2M} \bm\gamma}{\pi^{4M}} e^{ -2 (\bm\alpha_i - \bm\beta )^{\dag}(\bm{s'}_i - \bm{s}_i)^{-1}(\bm\alpha_i - \bm\beta )}\\\times e^{ -2 (\bm\alpha_{i+1} - \bm\gamma )^{\dag}(\bm{s}_{i+1} - \bm{s'}_{i+1})^{-1}(\bm\alpha_{i+1} - \bm\gamma )} \\\times\Lambda_{\mathcal{E}_i}^{(-\bm{s}_i, \bm{s}_{i+1})}(\bm\beta,\bm\gamma).
\end{multline}
Since we are convoluting with a Gaussian, the function gets smoother and negative regions become shallower (see Eq.~\eqref{eq:Wttot}). Due to the sign difference, the same happens for $\bm{s'}_{i+1} \prec \bm{s}_{i+1}$. In other words, the phase-space representation $\Lambda_{\mathcal{E}_i}^{(-\bm{s}_i, \bm{s}_{i+1})}$ behaves more and more like a positive kernel as we increase $\bm{s}_i$ and decrease $\bm{s}_{i+1}$. 

At the $i$th step in the for-loop of Algorithm \ref{alg:main}, the input ordering parameters $\bm{\tau}_i$ are fixed in the previous step, $i-1$, to be as high as possible. As implied by Eq.~\eqref{eq:Conv}, this also makes $\Lambda_{\mathcal{E}_i}^{(-\bm{\tau}_i, \bm{s}_{i+1})}$ as close to being regular and positive as possible, therefore we shall now focus just on the effect of the output parameters $\bm{s}_{i+1}$ to show the consistency of our algorithm. 

This maximal value will then become $\bm{\tau}_{i+1}$. By virtue of Eq.~\eqref{eq:Conv}, the procedure of maximising $\bm{s}_{i+1}$ in step $i$ of Algorithm \ref{alg:main} guarantees the best possible choice for the first ordering
parameters in the transfer function of step $i+1$. Nevertheless, at any stage of the for-loop, it is possible that no appropriate ordering parameter can be found to guarantee that the transfer function is a positive kernel. In that case, the algorithm fails, and we cannot simulate \eqref{proba_distrib_sequence} through sampling in this manner, one would need to find another way to compile the gates, as it is discussed in the outlook.

Let us note that the concept of 'maximizing a vector' is ambiguous. Therefore, in the context of multimode gates, where one is confronted with ambiguity, we sought to maximize the minimal value of the vector. One will find an illustration, and a discussion on this issue, in Section \ref{sec:bs}, where the beam-splitter is studied. The order in which the algorithm is applied can be impacted by this ambiguity. Initiating the process in reverse, beginning with the measurement, has the potential to yield divergent results when such a gate is encountered. Consequently, this can result in different constraints being imposed on the $(s)$-parameter at the last step.

If we do make it all the way to the final step, we find a fixed parameter $\bm{\tau}_{k+1}$ for which we have to evaluate the POVM elements ${W_{\hat{\Pi}_{\bm{x}}}}^{(-\bm{\tau}_{k+1})}(\bm{\alpha}_{k+1})$. The function ${W_{\hat{\Pi}_{\bm{x}}}}^{(-\bm{\tau}_{k+1})}$ is a product of single-mode functions $W_{\hat{\Pi}_{x}}^{(-\tau^{(j)}_{k+1})}$, with $\int_{x \in \mathcal{X}_j} dx \, \hat{\Pi}_{x} = \identity$. The set $\mathcal{X}_j$ denotes the set of all possible measurement outcomes for the detector that measures the $j$th mode. This implies that every local detector is described by a set of phase-space representations $\{ W_{\hat{\Pi}_x}^{(-s)}(\alpha) \mid x \in \mathcal{X}_j \}$ that satisfy the condition
\begin{equation}
\pi \int_{x \in \mathcal{X}_j} d x\,  W_{\hat{\Pi}_x}^{(-\tau^{(j)}_{k+1})}(\alpha) = 1,
\end{equation}
which must hold for all $\alpha \in \mathbb{C}$. For the simulation algorithm to work, we must be able to sample a final measurement outcome $\bm{x}$ that describes the click in all the detectors. When the resulting parameter $\bm{\tau}_{k+1}$ is such that the functions $W_{\hat{\Pi}_x}^{(-\tau^{(j)}_{k+1})}(\alpha)$ are positive functions for all possible measurement $x\in \mathcal{X}_j$, for all the modes $j$, we can treat the functions $\pi W_{\hat{\Pi}_x}^{(-\tau^{(j)}_{k+1})}(\alpha)$ as probability distributions on the space $x\in \mathcal{X}_j$. If this is the case, we can sample the measurement outcomes for all the individual detectors and finally obtain $\bm x$ in a way that is consistent with probability density \eqref{proba_distrib_sequence}. \\

\begin{algorithm}
\caption{ Classical simulation algorithm for a quantum computational scheme following the general structure of Figure (\ref{fig:inter}). Notice that in the last step $\bm{\beta_{k+1}}$ is fixed and $ \pi^M W_{\Pi_{\bm x}}^{(-\bm\tau_{k+1})}(\bm{\beta_{k+1}})$ is a PDF in $\bm{x}$. }
\label{alg:main}
    \SetKwInOut{Input}{input}
    \SetKwInOut{Output}{output}
    \Input{set of functions $W_{\Rh_{in}}^{(\bm{s}_1)}, \{\Lambda_{\mathcal{E}_k}^{(-\bm{s}_i, \bm{s}_{i+1})}\}_{i\in[1,k]}, W_{\Pi_{\bm x}}^{(-\bm{s}_{k+1})}$}
    \Output{\textit{a sample of measurement outcomes according to $p(\hat{x})$} (result of classical simulation), \textit{failed} (classical simulation not possible)}
    \SetKwBlock{Beginn}{beginn}{ende} 
    \Begin{
    \Comment{\textit{First the optimization}} \\
    find the max $\bm{s}_1$ s.t. $W_{\Rh_{in}}^{(\bm{s}_1)}(\bm{\alpha_1})$ is non-negative \;
    let $\bm\tau_1 := \max(\bm{s}_1)$ \;
    \For{$i \in [1,k]$}{
        find max $\bm{s}_{i+1}$ s.t. $\Lambda_{\mathcal{E}_i}^{(-\bm\tau_i, \bm{s}_{i+1})}(\bm{\alpha_{i}},\bm{\alpha_{i+1}})$ is non-negative  \;
        \If{no such $\bm{s}_{i+1}$ can be found}{
           return: \textit{failed} \;
            } 
        \Else{
        let $\bm\tau_{i+1} := \max(\bm{s}_{i+1})$ \;
        }
    }
    \If{$W_{\Pi_{\bm x}}^{(-\bm\tau_{k+1})}(\bm{\alpha_{k+1}})$ is non-negative for $\bm\tau_{k+1}$}{
    proceed to sampling \;
    }
    \Else{return: \textit{failed} \;}
    \Comment{\textit{Then the sampling}} \\
    sample a point $\bm{\alpha_1}$ according to $W_{\Rh_{in}}^{(\bm\tau_1)} (\bm{\alpha_{1}})$ \;
    $\bm{\beta_1} := \bm{\alpha_1}$;\\
    \For{$i \in [1,k]$}{
        sample a point $\bm{\alpha_{i+1}}$ according to $\Lambda_{\mathcal{E}_i}^{(-\bm\tau_i, \bm\tau_{i+1})}(\bm{\beta_{i}},\bm{\alpha_{i+1}})$ \;
        $\bm{\beta_{i+1}} := \bm{\alpha_{i+1}}$ \; 
        }
    {sample a string of outcomes $\bm{x}$ according to $\pi^M W_{\Pi_{\bm x}}^{(-\bm\tau_{k+1})}(\bm{\beta_{k+1}})$ \;
    return: ${\bm x}$ \;
    }
}  
\end{algorithm}
The previously described optimisation scheme for the ordering parameters explains the foundations of Algorithm \ref{alg:main}. This algorithm effectively reduces the problem of the positivity of the quasiprobability distributions to a study of the ordering parameters. We are only concerned with those parameters and how they change during the computation. Indeed, Algorithm \ref{alg:main} fixes ordering parameters layer by layer, and we can understand this as an evolution of the initial ordering parameters (and thus nonclassical depth) of the input state throughout the different stages of the algorithm. Once the optimization part of the algorithm has identified the optimal ordering parameters, the second part of the algorithm, doing the sampling, can proceed in the same way as the one in Ref.~\cite{mari2012positive}. Considerations on computational complexity of the sampling algorithm were already provided there, whereas the focus of our work is on the characterization of the gates transfer functions in terms of the ordering parameters and on their optimal choice.

We emphasise that when the optimisation phase succeeds, the second phase of Algorithm 1 generates full measurement outcome strings $\bm{x}$ by sequentially sampling from the certified positive probability density functions, following a Markov-chain strategy. This constitutes a bona fide classical simulation of the sampling problem, producing outcome strings distributed according to $p(\mathbf{x})$, and not merely an estimation of individual outcome probabilities.

The pseudo-algorithm presented here is intended to serve as a conceptual framework, to which mathematical considerations will be applied in the following sections. Nevertheless, the interested reader may refer to the explicit code provided \cite{Frigerio_s_-QPDs_optimization_tool} to check simulatability for arbitrary photonic quantum circuits involving a universal set of gates, specifically the one characterized in Section \ref{sec5}. Given such a circuit and the non-classical depths of the (multimode, factorized) input states, the code provides two possible outcomes: either the quest for positivity conditions fails at some point in the computation, or succeeds until the end. In the first instance, the code outputs the gate whose transfer function could not be made positive. In the latter case, the user is asked to input the nonclassical depths associated with the measurements POVM and the code checks the classical simulatability of the full scheme by comparing the latter with the final optimized $s$ parameters. We emphasise that, for the universal gate set characterised in Sec. 5, the maximisation of the output ordering parameter at each step reduces to evaluating closed-form inequalities (Eqs. (25), (27)–(29), and Theorem 2), rather than requiring a generic numerical optimisation. The code \cite{Frigerio_s_-QPDs_optimization_tool} implements these analytic conditions and chains them sequentially through a user-specified circuit.

We should emphasise some key assumptions from the computational point of view. First of all, we assume that the circuit decomposition $\mathcal{E} = \mathcal{E}_1 \circ \mathcal{E}_2 \circ \dots \circ \mathcal{E}_k$ is known. Furthermore, we assume that for all the elements in the circuit (input states, gates, and measurements) the $(s)$-parameterised phase space representations in \eqref{proba_distrib_sequence} are known (or can be calculated at a negligible overhead). Finally, we assume that it is computationally efficient to sample from these PDFs once they are positive. This final assumption may break down once the functions oscillate too rapidly, an effect which is in principle captured by the stellar rank \cite{PhysRevLett.130.090602}.

To encapsulate, in broad terms, the fundamental principle underlying the present method we propose the following simplified thought experiment. Let us assume that a single-mode highly non-classical state goes through a specific channel whose transfer function, $\Lambda_{\mathcal{E}}^{(-\tau_{in}, \tau_{out})}$, is positive only for $\tau_{out}$ arbitrarily close to $1$. The inherent characteristic of the input state would force the first parameter to assume a value $\tau_{in}$ close to $-1$. Indeed, if one whishes to erase all negativity from the Wigner function of such a state, one would have to proceed all the way to the smooth, completely positive Husimi $Q$-function by convolving said Wigner function with a broad Gaussian. As such, the channel has effectively transformed an input state that is highly non-classical into an output state that is highly classical; a greater range of quasi-PDFs behave as PDFs at the output than at the input. Such a channel would represent a point in the computation where all potential quantum advantage that might have been provided by the nonclassical input is effectively lost. This channel is ineffective in terms of providing a quantum advantages, as it effectively negates the potential benefits of imputing a Wigner-negative state. This highlights that it is crucial to understand the relation between the parameters $\tau_i$ and $\tau_{i+1}$ that are selected by Algorithm \ref{alg:main}. Therefore, in the next section, we will examine different elementary gates and observe how the ordering parameter evolves through them.

\section{Resourcefulness of Gates} \label{sec:Gates}
\subsection{Conditions for a general unitary}
Before delving into less trivial examples, important conclusions on the ordering parameters can be drawn from the study of the identity channel. Since it acts independently on each mode, we will detail it for a single mode. Moreover, given that this channel leaves any input state unchanged, it cannot modify the nonclassical depths. Consider then an input state $\Rh$ such that $W^{(s_i)}_{\Rh}$ is a regular probability density function and $s_i$ is the largest possible value, i.e.  corresponding to the nonclassical depth. If the transition kernel $\Lambda^{(-s_{i}, s_{i+1})}_{\identity}$ were to be positive and regular for $s_{i+1} > s_i$, we would obtain as an output a function $W^{(s_{i+1})}_{\Rh}$ which is still regular and positive, contradicting the hypothesis that $s_i$ was the largest value. This implies that a necessary condition for $\Lambda^{(-s_{i}, s_{i+1})}_{\identity}$ to be regular and positive is that $s_{i+1} \leq s_i$, which is also a sufficient condition by Eq. (\ref{conditionCaves}) with $L = \identity$ being the identity matrix. When this condition is met $\Lambda^{(-s_{i}, s_{i+1})}_{\identity}$ is a Gaussian function in its input and output variables, while for $s_{i+1} > s_i$ it is a singular distribution attaining divergent values almost everywhere.

Likewise, it can be foreseen that for a single-mode unitary channel $\hat{U}$ the condition $s_{i+1} \leq s_i$ is still necessary in order for $\Lambda^{(-s_{i}, s_{i+1})}_{\hat{U}}$ to be regular and positive, otherwise $\hat{U}$ could increase the nonclassical depth of \emph{any} input state with initial nonclassical depth $\mathfrak{t}$. Let us prove this formally:

\begin{thm}
    Let $\hat{U}$ be a single-mode unitary operator. If the associated $(-s_i,s_{i+1})$-ordered transfer function $\Lambda^{(-s_{i}, s_{i+1})}_{\hat{U}}$ is a regular function, then $s_{i+1} \leq s_i$. 
\end{thm}

\begin{proof}
 Consider the equality $\hat{U} \hat{U}^{\dagger} = \identity$ rewritten in terms of $s$-ordered quasi-PDF:
\begin{eqnarray}
 \label{eq:proofU}
   \int_{\alpha_{2} \in \mathbb{C}} d^{2} \alpha_{2} \  \Lambda^{(-s_{i}, s_{i+1})}_{\hat{U}} ( \alpha_{1}, \alpha_{2} ) && \Lambda^{(-s_{i+1},s_{i+1})}_{\hat{U}^{\dagger}} (\alpha_{2} ,\alpha_{3} ) \ \nonumber = \\
    &&  \ \Lambda^{(-s_{i}, s_{i+1})}_{\identity} ( \alpha_{1}, \alpha_{3} ).
\end{eqnarray}
Formally, we can write the right-hand side as:
\begin{eqnarray}
    \Lambda^{(-s_{i}, s_{i+1})}_{\identity} ( \alpha_{1}, \alpha_{3} ) \ \propto \ \int_{\xi \in \mathbb{C}} d^2 \xi \ && e^{ (\alpha_{1} - \alpha_{3}) \xi^\dagger -  (\alpha_{1} - \alpha_{3})^\dagger \xi} \nonumber\\
    && \times e^{ \xi (s_{i+1}-s_i) \xi^\dagger /2}.
 \end{eqnarray}
If $s_{i+1} > s_{i}$, the former expression diverges almost everywhere in $\alpha_{1}, \alpha_{3}$ and it is more singular than a delta distribution, while for $s_{i+1} = s_{i}$ it reduces to the delta distribution, imposing $\alpha_{1} = \alpha_{3}$. Now if $\Lambda^{(-s_{i}, s_{i+1})}_{\hat{U}}$ were to be a regular, positive semidefinite function \footnote{We shall assume that whenever an $s$-ordered quasiprobability is regular and non-negative, it also belongs to the Schwartz space of rapidly decreasing function, so that no issues of convergence can arise with respect to tempered distributions. This is typically the case, since this probability density functions decay as Gaussians for large absolute values of their arguments.}, the result of the left-hand side in Eq. (\ref{eq:proofU}) cannot be divergent, even if $\Lambda^{(-s_{i+1},s_{i+1})}_{\hat{U}^{\dagger}} $ is a very singular distribution, which implies that indeed $\Lambda^{(-s_{i}, s_{i+1})}_{\hat{U}}$ can never be a regular probability density function if $s_{i+1} > s_i$. Notice that when $s_{i+1} = s_i$ we could have, in the best case, that $\Lambda^{(-s_{i}, s_{i+1})}_{\hat{U}}$ is as singular as a delta distribution and then Eq. (\ref{eq:proofU}) still holds even if the right-hand side diverges for $\alpha_{1} = \alpha_{3}$. 
\end{proof}

\subsection{Losses}
Before providing the characterization of a universal set of unitary quantum gates, it is instructive to contrast the general result that we got for unitary channels with the prototypical example of dissipative quantum channels, i.e. losses which are crucial in modelling optical setups and deserves an additional discussion in its own right. The optical loss channel is implicitly already covered in Ref.~\cite{rahimi2016sufficient} where Eq.~\eqref{conditionCaves} is derived to describe a linear optics network with losses. To consider optical losses in the spirit of our gate-based protocol of Fig.~\ref{fig:inter}, we can just define a single-mode loss channel $\Lambda_{\mathcal{E}_{\eta}}^{(-s_{i},s_{i+1})}(\alpha,\beta)$ as a single-mode version of Eq.~\eqref{conditionCaves} setting $L = \sqrt{\eta}$, where $\eta \in [0,1]$ describes the efficiency of the channel ($\eta = 1$ reflects the absence of losses, while $\eta = 0$ means that the state is completely lost). We can from this find a decreasing linear relation with respect to $s_i$ :
\begin{equation}
   \label{losses_result}
    s_{i+1} \geq 1 - \eta (1-s_i).
\end{equation}
Since $s_i =( 1-\eta) s_i + \eta s_i <  1 - \eta (1-s_i)$, this implies that we can always pick $s_{i+1} > s_i$ for $\eta < 1$ and, the smaller the value of $\eta$, the closest the maximum value of $s_{i+1}$ approaches $1$, as shown in Fig.~\ref{fig:losses-one}. It can thus be concluded that losses always result in a more classical output, in stark contrast with the opposite result that we derived for unitary gates \ref{eq:proofU}, thereby emphasizing the peculiar nature of losses for nonclassicality. This relatively straightforward result, moreover, allows us to understand how losses can hinder a possible quantum computational advantage. In a real circuit, loss channels of the type $\Lambda_{\mathcal{E}_{\eta}}^{(s_{i},s_{i+1})}$ can occur everywhere, effectively increasing the chances of success of Algorithm \ref{alg:main} at any step. As is commonly done in quantum optics, we can now describe realistic operations and measurements by taking ideal unitary ones and adding a loss channel at the input and output. When we add them in between the gates and run Algorithm \ref{alg:main}, we can now deal with mode-dependent losses, which are often overlooked in studies of sampling problems. 

Finally, it is worth mentioning that losses allow us to consider in principle values of the $s$ parameters outside the range $[-1,1]$: indeed a coherent state going through a loss channel will become a displaced thermal state, whose $s$-ordered quasi-PDF will be a regular and positive function also for some values of $s>1$, defined again through Eq. (\ref{eq:sParameterDef}). In general, problems might arise because the $s$-ordered quasi-PDFs of channels and states for values of $s>1$ are not guaranteed to exist, even in the sense of distributions, and they are not associated with an ordering of the creation and annihilation operators. However, if we have input states that are mixed and noisy, such that they \emph{can} be parametrized with values of $s$ larger than $1$, the matching rule in Eq. (\ref{eq:swithminuss}) ensures that at the next step we will have an even more regular function (since $s$ matches with $-s$) and we can possibly have more chances to find a positive decomposition of the whole circuit by going beyond the limitation of $s \in [-1,1]$.

\section{Characterization of a Universal Set of Unitary Gates}
\label{sec5}
Algorithm \ref{alg:main} is based on the idea that we have at hand a decomposition of the full photonic quantum circuit into a set of previously characterized gates, such that the computational effort needed to compute the $(s)$-ordered quasi-PDFs for such elements constitutes just a fixed preliminary overhead with respect to the actual simulation cost of the computation through sampling. It is thus consequential to characterize a universal set of unitary gates, with which all CV unitary operations can be approximated efficiently \cite{PhysRevLett.82.1784}; one such set is provided by \emph{two-mode Gaussian unitary gates} and one single-mode non-Gaussian gate, such as \emph{the cubic phase gate} \cite{Hillmann2020}, whose characterizations are provided in the following.
\subsection{Gaussian unitaries}
Among Gaussian unitary gates, displacements and single-mode phase shifters clearly do not affect the nonclassicality (hence they leave the value of the $s$-parameters unchanged), since they act as Euclidean transformations on quasi-PDFs. Thus, we only need to consider squeezers and beam splitters. We will show that the action of a squeezing gate can directly alter the $s$ parameter of the output, whereas beam splitters can ``mix'' the $s$ parameters of their input modes.

\subsubsection{Squeezing gate} \label{squeezing}
\begin{figure*}[hbt]
\centering
\subfigure[]{\includegraphics[height=5.25cm]{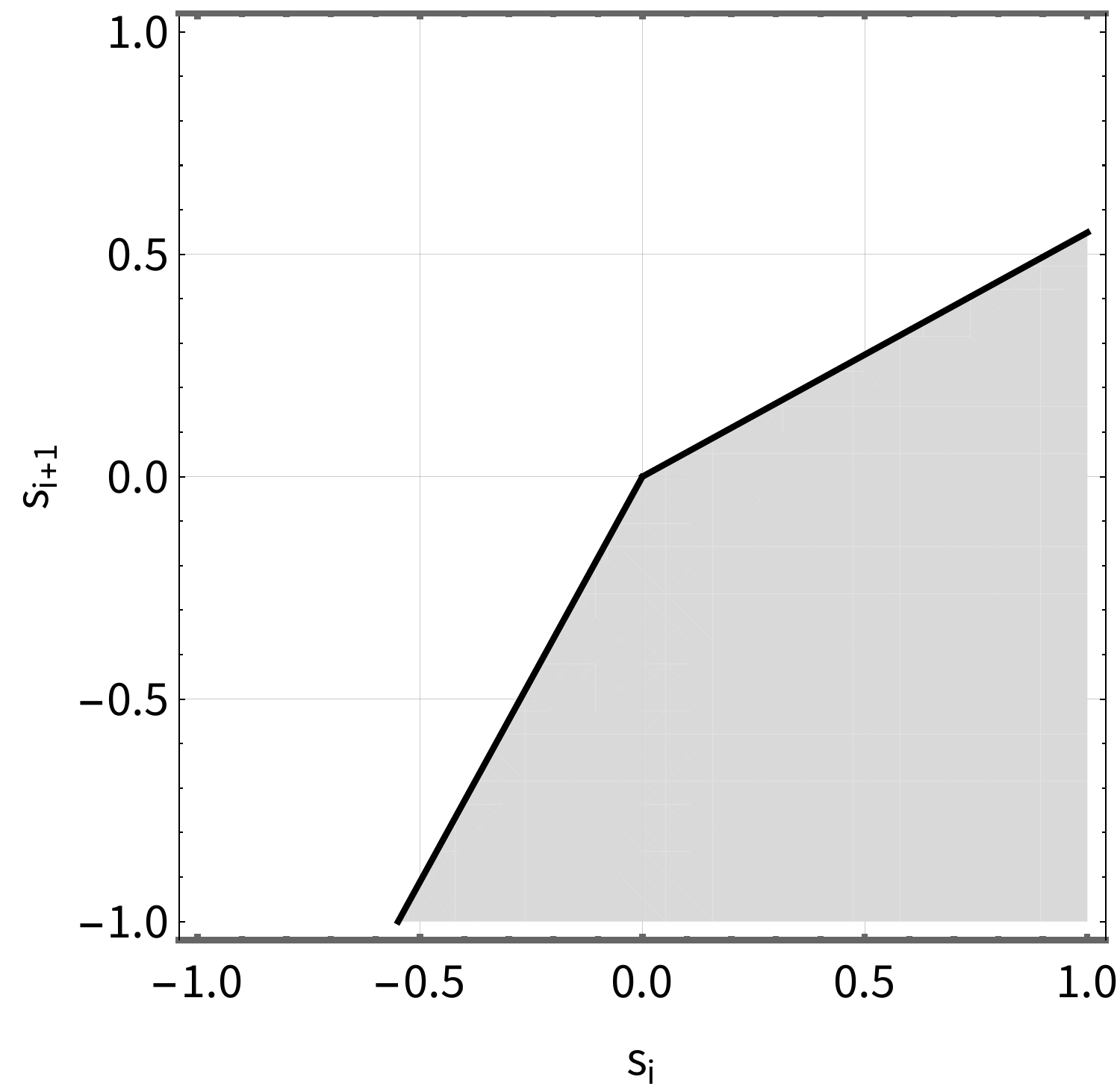}
    \label{fig:squeezing} }
\subfigure[]{\includegraphics[scale=0.25]{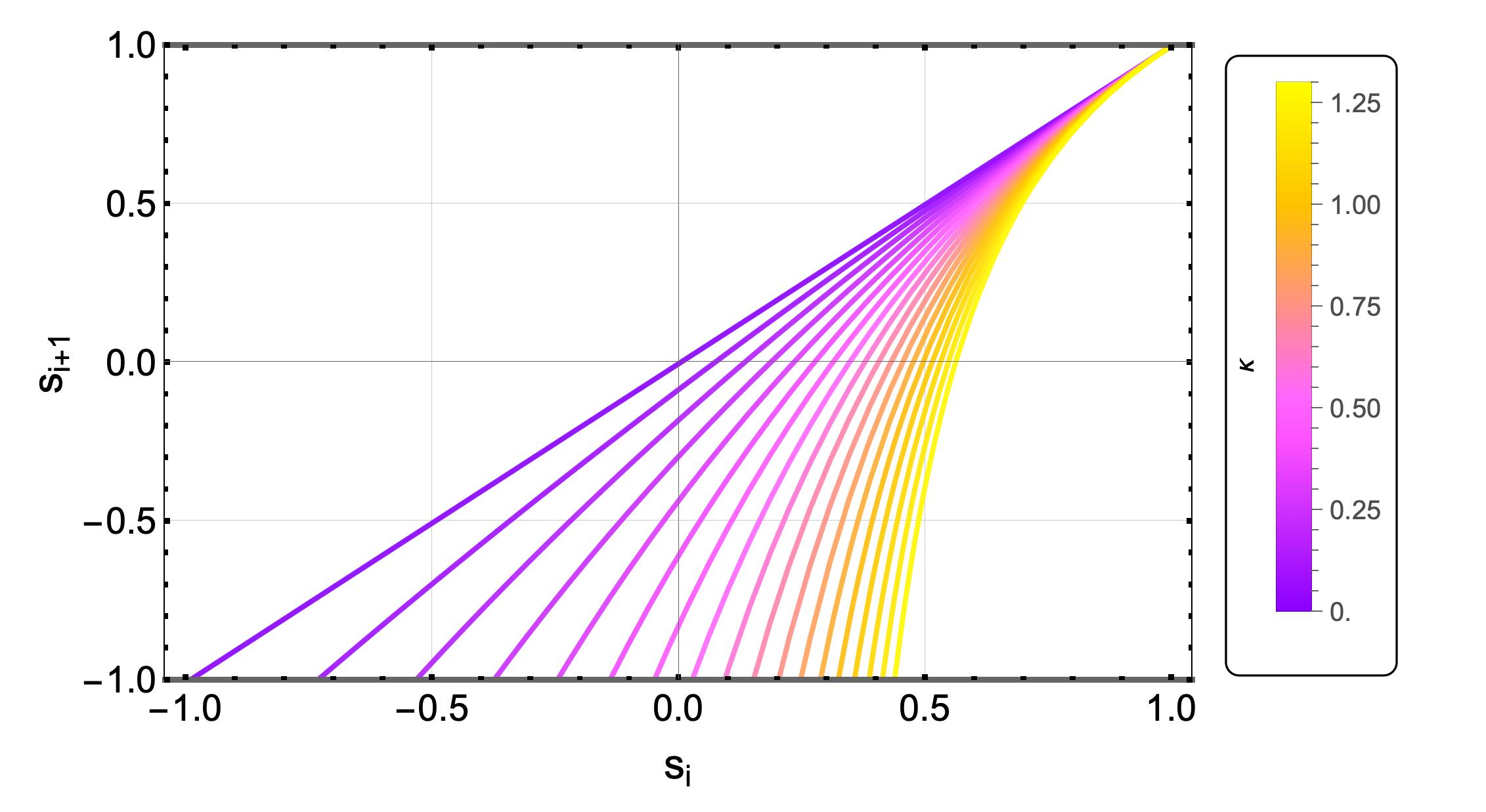} \label{fig:photsubtrparam}}
\caption{(a) On the squeezing gate: the region in the $(s_{i},s_{i+1})$ plane where $\Lambda_{\mathcal{S}}^{(-s_i,s_{i+1})}$ is non-negative according to Eq. (\ref{conditions_squeezing}) is displayed in gray, for $r = 0.3$. (b) On the single-photon subtraction: limit lines according to Eq. (\ref{eq:photsubtrcond}) for various values of $\alpha \in [0,1]$, represented by different shadings. For each value of $\alpha$, the allowed values of $s_{i+1}$ to have a non-negative transfer functions of the BS lie under the respective curve.}
\end{figure*}

Starting with the squeezing gate, denoted by $\mathcal{S}$, the trace-preserving quantum gate $\mathcal{S}$ acting on the displacement operator is:
\begin{equation}
    \mathcal{S}(\hat{D}^\dag(\xi)) = \hat{S}^\dag(r)\hat{D}^\dag(\xi)\hat{S}(r) = \hat{D}^\dag(\chi),
\end{equation}
where $\hat{S}(r)$ is the squeezing operator for a squeezing parameter $r$ and $\chi = \xi\cosh r+\xi^\dag \sinh r$. Thus implementing a simple re-scaling of the displacement operator. For $\mathcal{S}$ the transfer function take the form 
\begin{multline}
    \label{lambda_squee}
    \Lambda_{\mathcal{S}}^{(s_i,s_{i+1})}(\alpha,\beta) \\ = \iint \frac{d^2\zeta}{\pi^2}\frac{d^2\xi}{\pi} e^{\beta \zeta^\dag - \zeta\beta^\dag} e^{\zeta s_{i+1}\zeta^\dag/2} e^{\xi\alpha^\dag - \alpha \xi^\dag} e^{\xi s_i\xi^\dag/2} \\ \times \Tr[\mathcal{S}(\hat{D}^\dag(\xi))\hat{D}(\zeta)].
\end{multline}
We can write 
\begin{equation}
    \mathcal{S}(\hat{D}^\dag(\xi)) = e^{\chi \chi^\dag} \int \frac{d^2\gamma}{\pi} e^{\gamma \xi^\dag - \xi \gamma^\dag} \ket{\gamma}\bra{\gamma}.
\end{equation}
And we can use the normally ordered form of the displacement operator, $\hat{D}(\zeta) = e^{-\zeta\zeta^\dag/2}e^{\zeta\hat{a}^\dag}e^{-\hat{a}\zeta^\dag}$ to obtain
\begin{equation}
    \Tr[\mathcal{S}(\hat{D}^\dag(\bm{\xi}))\hat{D}(\bm{\zeta})] = \pi e^{\chi \chi^\dag/2}e^{\zeta \zeta^\dag/2}\delta(\zeta-\chi) 
\end{equation}
Thus the transfer function (\ref{lambda_squee}) becomes
\begin{equation}
    \Lambda_{\mathcal{S}}^{(-s_i,s_{i+1})}(\alpha,\beta) = \int d\xi e^{\chi^\dag\beta-\chi\beta^\dag+\frac{\chi s_{i+1}\chi^\dag}{2}} e^{\xi^\dag\alpha-\xi\alpha^\dag+\frac{\xi s_i\xi^\dag}{2}}
\end{equation}
As one can notice from this form the transfer function is well behave and non-negative if and only if
\begin{eqnarray}
    \sigma^{-1} =&& [(\xi\cosh r+\xi^\dag \sinh r) s_{i+1}(\xi^\dag\cosh r+\xi \sinh r) \nonumber\\
    &&- s_i|\xi|^2]^{-1} \ge 0
\end{eqnarray}
Finally, we find that one can sample from the transfer function, and thus classically simulate this block if and only if 
\begin{equation}
    \label{conditions_squeezing}
    \left\{
        \begin{array}{ll}
            s_{i+1} \le \dfrac{s_i}{e^{2|r|}} \;\;\; \mbox{if } s_i\ge0\\
            s_{i+1} \le \dfrac{s_i}{e^{-2|r|}} \;\;\; \mbox{if } s_i\le0
        \end{array}
    \right.
\end{equation}
Therefore, in general, the squeezing channel requires us to parametrize the output expecting it to have a greater nonclassical depth (lower maximum value allowed for $s$) in order for the transfer function to be a positive transfer probability kernel. It is interesting to note that the conditions linearly bound one ordering parameter with respect to the other and they are fully symmetric under the exchange $-s_{i} \leftrightarrow s_{i+1}$, as can be seen more clearly from Fig. (\ref{fig:squeezing}), where we plot in gray the region in the $(s_{i},s_{i+1})$ plane where $\Lambda_{\mathcal{S}}^{(-s_i,s_{i+1})}$ is non-negative, for $r = 0.3$. 

Notice that, for $s_{i} < 0$, the upper bound $s_{i} / e^{-2\vert r \vert}$ can be smaller than $-1$, namely if $\vert r \vert > - \frac{1}{2} \ln{ \vert s_{i} \vert }$, therefore for sufficiently nonclassical, Wigner negative input states the squeezing channel cannot be simulated through Algorithm 1, no matter how the output is parametrized. This observation provides a first hint of the fact that non-Gaussian resources are pivotal for possible quantum advantages.

It is crucial to stress that one could include the squeezing channel into the previous channel that was applied to the multi-mode system and try to simulate it as a single block with a positive transfer probability kernel, possibly finding less strict conditions for the output parameter $s_{i+1}$; however, when the previous channel is multi-mode and potentially entangling, this forces us to apply the squeezing to a multi-mode state, converting the challenge of simulating the possibly negative squeezing transfer function into that of simulating a positive transfer function on a larger space.

A special point deserves now our attention, the output state of the squeezing channel may not be more non-classical than the input. It is evident that the conditions set out in Eq. (\ref{conditions_squeezing}) pertain to the positivity and well-behaveness of the transfer function. In the event that the aforementioned conditions are not met, the transfer function will exhibit negative region(s) and/or will be highly singular, thereby implying that one cannot sample from it efficiently \cite{pashayan2015estimating}. Nevertheless, it is possible for the output state to be, in all respects and according to all characterisations, classical. This can be understood by immagining a highly squeezed input state going through a squeezing channel with the opposite squeezing parameter and resulting in a more classical output. Through the action of $\mathcal{S}$, we then squeeze in the orthogonal direction, by the same factor of $r$. In this case, the output state would be very classical. However, if the conditions set out in (\ref{conditions_squeezing}) are not met, it is not possible to sample from the transfer function describing the process.

\subsubsection{Beam-splitter}
\label{sec:bs}
We shall now proceed to identify the conditions under which the beam-splitter (BS) can be considered a resourceful gate. Assume that we encounter a BS gate on block $i$, where $1< i \le k$. In a formal sense, our objective is to construct a channel $\mathcal{E}^{BS}$ that acts as a beam-splitter transformation between two modes and trivially on the remaining $M-2$ modes. Denoting by $a$ and $b$ the two modes on which the BS acts non-trivially, according to Eq. (\ref{conditionCaves}) the limiting condition relating output to input ordering parameters is obtained by taking $L$ is a unitary matrix describing the beam-splitter transformation,
\begin{equation}
    L = \begin{pmatrix}
\cos \theta  & \sin \theta \\
- \sin \theta  & \cos \theta 
\end{pmatrix},
\end{equation}
and $\theta \in \mathbb{R}$. Potential phase factors $e^{i\phi}$, representing the phase shift between the input modes of the beam splitter, has not been taken into account since the phase shift channel has no impact on the ordering parameter, merely rotating phase-space distributions without altering their values. By virtue of unitarity, one needs to solve the following equation maximizing the output parameters:
\begin{equation}
    \label{bm_eq}
   \min {\rm eig}\left( - \diag(s_{i+1}^a, s_{i+1}^b) + L^\dag \diag(s_{i}^a, s_{i}^b)L \right)= 0.
\end{equation}
While the relation between the input and output parameters is not very insightful at first glance, we do see that $s_{i+1}^a$ will generally depend on both $s_{i}^a$ and $s_{i}^b$ (and analogously for $s_{i+1}^b$). Accordingly, the process of maximizing the output $(s)$-parameters for a given input is not well defined and a continuous range of solutions may exists, depending on the relative balance in the combined maximization. To keep consistency with the proposed algorithm we impose the symmetric condition $s_{i+1}^a=s_{i+1}^b$, which is equivalent to maximizing the minimum of the two. It is useful to notice that from Eq. (\ref{bm_eq}) one can derive a necessary condition on the sum of the output $s$-parameters:
\begin{equation}
    s^{a}_{i+1} + s^{b}_{i+1} \ \leq \ s^{a}_{i} + s^{b}_{i} - \vert s^{a}_{i} - s^{b}_{i} \vert  \sin 2 \theta 
\end{equation} 
assuming $0 \leq \theta \leq \frac{\pi}{2}$. In particular, if one imposes $s_{i+1}^a=s_{i+1}^b$ for a balanced BS ($\theta = \pi/4$), one gets the necessary and sufficient condition for the positivity of the transfer function:
\begin{equation}
    s_{i+1}^a=s_{i+1}^b \leq \min \{ s^{a}_{i}, s^{b}_{i} \} 
\end{equation}
Thus the most nonclassical of the two inputs fully dictates the maximum value for the output ordering parameters.

\subsection{Realistic protocol for single-photon subtraction}\label{subtraction}
\begin{figure}[H]
    \centering
    \includegraphics[width=0.7\linewidth]{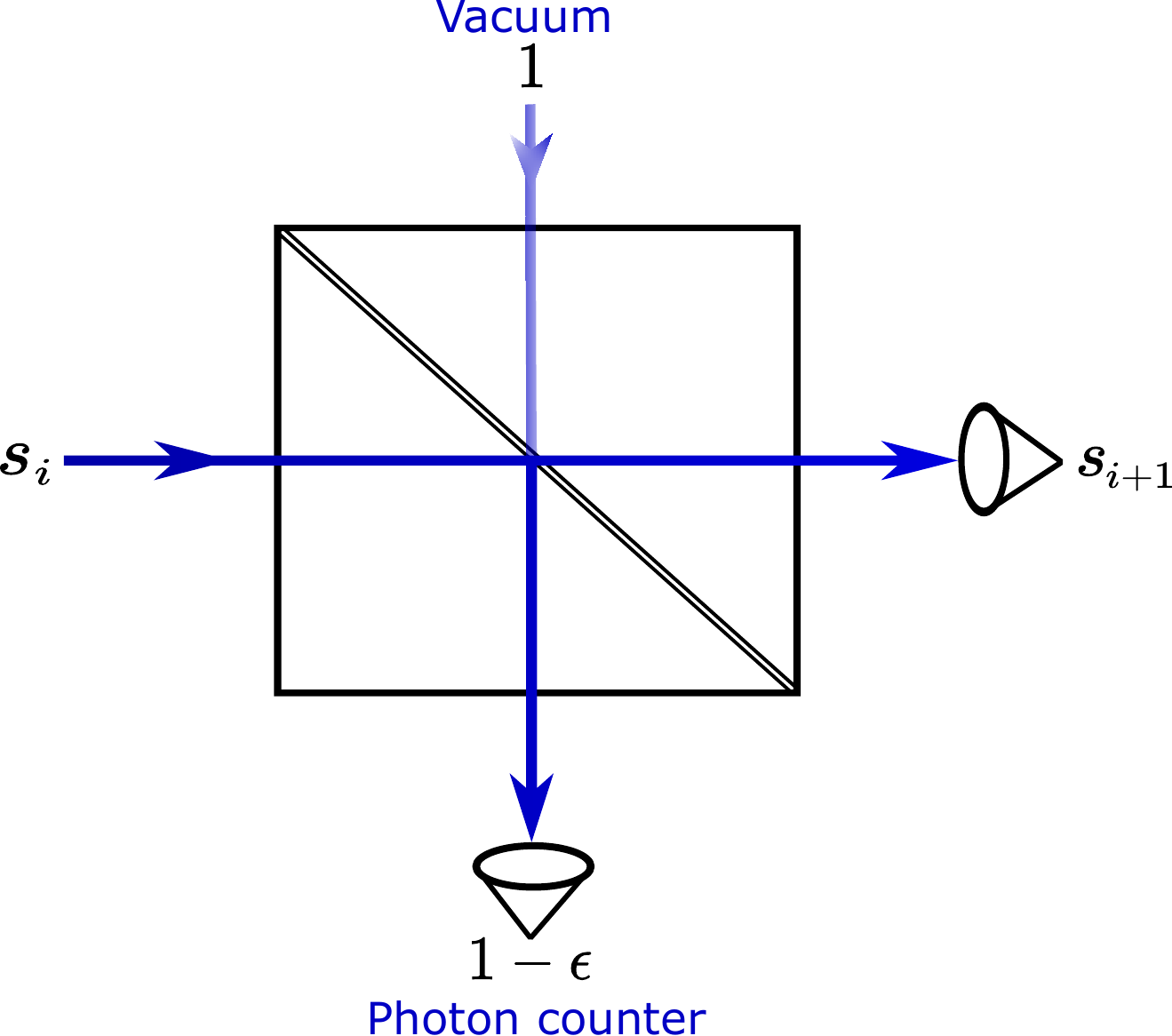}
    \caption{Realistic single-photon subtraction protocol seen from the perspective of our framework, where we considers solely the order parameter of the states on each branch.}
    \label{fig:bs_subtraction}
\end{figure}
In order to showcase the usefulness of the proposed protocol, we will apply it to the single-photon subtraction operation which is commonly considered as one of the most accessible de-Gaussifying procedures. We model realistic single-photon subtraction with a BS whose input modes are respectively the input state on which we wish to subtract a photon on one branch, and the vacuum state on the other. The transmittance $\eta = \cos^2 \theta$ of the BS with respect to the first mode is close to $1$, meaning that only a small fraction of the signal will be reflected. On this weak output signal, a photon counter with some finite efficiency will be implemented, corresponding to the ideal POVM ($\epsilon = 0$) with elements $\{ \hat{\Pi}_1 =  \ket{1}\bra{1}, \hat{\Pi}_0 = \identity-\ket{1}\bra{1}\}$.
At the last step, only the outcomes corresponding to $\hat{\Pi}_1$ are post-selected. This measurement will be modelled choosing an ordering parameter close to $-1$, since the only value such that the $s$-ordered quasi-PDF of $\ket{1}\bra{1}$ is positive is $-1$, being a projector on a highly non-classical state, but we shall assume that its value is $-1 + \epsilon$ to account for imperfect measurements. This proposed model is illustrated in Fig. \ref{fig:bs_subtraction}.

This results in the following set of input-output ordering parameters:
\begin{equation}
    \bm{s}_{in} = \begin{pmatrix}
    s_i & 0\\
    0 & 1
    \end{pmatrix}; \;\;\; \bm{s}_{out} = \begin{pmatrix}
    s_{i+1} & 0\\
    0 & 1-\epsilon
    \end{pmatrix}
\end{equation}
recalling that the output parameter for the mode that is measured has the opposite sign with respect to the ordering parameter that we had to choose for the highly nonclassical measurement.
The resulting output parameter, $s_{i+1}$, with which we have to parameterize the output to have a positive transfer function for the beam-splitter will depend on the only free parameter : the ordering parameter of the input state.
Clearly, if we take the limit $\theta \to 0$ for finite $\epsilon$ the result will be trivial, having $s_{i+1} \leq s_{i}$ as the only condition. Going back to Eq. (\ref{bm_eq}), inserting the diagonal matrices above and imposing that the smallest eigenvalue is non-negative, for an ideal detection ($\epsilon = 0$) with generic mixing angle $\theta$ we find instead: 
\begin{widetext}
\begin{equation}
    \lambda_{min} \ = \ \dfrac{1}{2} \left[ s_{i} - 
   s_{i+1}  - \sqrt{(1-s_{i})^2 + (1-s_{i+1})^2 -2 (1 -s_{i}) (1-s_{i+1})\cos 2 \theta)} \right] \ \geq \ 0.
\end{equation}
\end{widetext}
For $-1 \leq s_{i+1} \leq s_{i} \leq 1$, this quantity is zero if $\theta$ is a multiple of $\pi$, including $\theta=0$, and it is negative otherwise. Therefore we conclude that there exists no set of input-output parameters such that the transfer function is well-behaved in the case of perfect photodetection. This result is of great importance as it demonstrates that ideal single-photon detectors are sufficient to render this building-block impossible to simulate classically. \\

There exists however a nontrivial limit, tuning in on the scaling $\theta \propto \sqrt{\epsilon}$.
In this case, taking the limit $\epsilon \to 0$ with $\theta = \kappa \sqrt{\epsilon}$ for a free tuning parameter $\kappa \in \mathbb{R}$, we can rewrite the condition $\lambda_{min} \geq 0$ as:
\begin{equation}
\label{eq:photsubtrcond}
    s_{i+1} \ \ \leq \ \ \dfrac{s_i - (1 - s_i) \kappa^2}{1 - (1 - s_i) \kappa^2}.
\end{equation}
In Fig. \ref{fig:photsubtrparam} these curves for various values of $\kappa$ are displayed, with smaller values corresponding to shades towards the violet. We notice that, for each value of $\kappa > 0$, there is a minimum $s_{i}$ under which no solution exist (each curve reaches the value $s_{i+1} = -1$ for some $s_{i}$ strictly greater than $-1$). Furthermore, for $\kappa > \frac{1}{2}$ this limiting value for $s_{i}$ is strictly positive, meaning that the single-photon subtraction modelled by this limit cannot be described with a non-negative transfer function whenever we input a Wigner-negative state. Significantly though, this implies that if we input noisy squeezed states, at fixed detection efficiency and transmittance, there is a maximum ratio of squeezing to noise beyond which the transfer function will never be positive, rendering this block non-simulatable even if the input is Wigner-positive.\\

Finally, let us note that the above arguments are purely based on the order parameters and do not require any further details of the phase-space representations. This means that this reasoning can be extended to virtually all other measurement-based operations.

\subsection{Non-Gaussian unitary gates}

Having seen that only Wigner-negative inputs can force the squeezing channel to be represented only by a negative transfer function, we will now tackle non-Gaussian single-mode unitary channels, which are needed to enable universal quantum computation. The key result that we obtained is the following:
\begin{thm}
\label{thm:nGunitaries}
    Let $\hat{U}$ be a single-mode unitary non-Gaussian gate. Then $\Lambda^{(s_1,s_2)}_{\hat{U}}(\alpha,\beta)$ cannot be an everywhere non-negative function for any $s_1,s_2 > -1$, while it is always non-negative for $s_1 = s_2 = -1$. 
\end{thm}
\begin{proof}
    Consider the application of $\hat{U}$ on coherent states. If each coherent state were mapped by $\hat{U}$ to some Gaussian state, it follows that every Gaussian state would be mapped to a Gaussian state via the Glauber representation theorem and the channel would be Gaussian. Since we supposed $\hat{U}$ to be non-Gaussian, there must then exist a coherent state $\vert \alpha_0 \rangle$ such that $ \vert \psi \rangle = \hat{U} \vert \alpha_0 \rangle$ is non-Gaussian. Since it is pure, non-Gaussianity implies that it has maximal nonclassical depth and only its Husimi Q-function is a proper probability distributions, while all the other $s$-ordered quasiprobability representations of $\vert \psi \rangle$ are non-positive for any $s > -1$. On the other hand, coherent states are $P$-classical and their Glauber $P$-distribution is a delta distribution, therefore we can choose $s_{in} = 1$ and $s_{1} = - s_{in} = -1$. Then we have:
    \begin{equation}
        \begin{aligned}
        W_{\vert \psi \rangle }^{(s_{2})}(\beta) \ & \propto \int \delta (\alpha - \alpha_0) \Lambda^{(-1,s_{2})}_{\hat{U}}(\alpha,\beta) d^2 \alpha \ = \\ & = \ \Lambda^{(-1,s_{2})}_{\hat{U}}(\alpha_0,\beta)
        \end{aligned}
    \end{equation}
    And since we showed that the left-hand side must attain negative values for some $\beta$ whenever $s_{2} > -1$, the first part of the thesis follows. Finally, notice that for $s_{1} = s_{2} = -1$ we have the very general result:
     \begin{equation}
       \Lambda^{(-1,-1)}_{\hat{U}}(\alpha,\beta) \propto \vert \langle \beta \vert \hat{U}\vert \alpha \rangle \vert ^2
 \end{equation}
 therefore the $(-1,-1)$-ordered transfer function is always non-negative. 
\end{proof}

\subsubsection{The cubic phase gate}
Having seen that only Wigner negative inputs can force the squeezing channel to be represented only by a negative transfer function, we can tackle an instance of a non-Gaussian single-mode unitary channel of utmost relevance in the theory of CV quantum computation: the cubic phase gate. It can be described by the unitary operator $\hat{U}_{\gamma} = e^{-i \frac{\gamma}{3!} \hat{q}^3}$, where $\gamma \in \mathbb{R}$ is known as the \emph{cubicity}. It is well known \cite{PhysRevLett.82.1784,GKP2001,gu_quantum_2009} that, together with Gaussian gates, the cubic gate is sufficient to attain universality with CV quantum computation. 

The explicit calculations are shown in Appendix (\ref{app:B}), where the $(s_1,s_2)$-ordered transfer function is given in two versions, one involving a single integral with a cubic phase term and another involving a double convolution of the Airy function with Gaussian kernels. This example also showcases how our method can be applied to the characterization of more complex unitary gates.


\section{Conclusions}

\subsection{Simulability of photonic circuits}
Using the language of $(s)$-ordered quasiprobability distributions on optical phase-space, we detailed a procedure to systematically assess the possibility of framing a computational photonic quantum process in terms of probability distributions updated through positive-semidefinite transition kernels. Whenever such a decomposition is possible, the process can be simulated efficiently by a classical algorithm, namely by sampling the starting probability distribution and applying the steps of the algorithm in order to compute the output. In doing so, we go beyond the already established necessity of Wigner negativity by enlarging the class of functions, allowing us to unveil bounds in which new quantum resources may be identified. In particular, our method is able to determine the exact amount of losses that are sufficient, at each step in the quantum computation, to make the corresponding process efficiently simulatable. We emphasize that the strength of our method heavily relies on the fact that the conditions for positivity of each transfer function are determined just by the ordering parameters, and therefore only by the nonclassical depth of the input and not on the specific quantum state. This property allows us to explicitly derive these positivity conditions for all Gaussian unitary channels, as well as for non-Gaussian unitary channels for ordering parameters in the range $[-1,1]$. In particular, a single-mode squeezing channel can always be implemented by a positive transfer function by some choice of the second ordering parameter, unless the input is Wigner negative. Studying how a beam splitter mixes the nonclassicalities of the input and fixing a maximally nonclassical measurement, we deduce that (realistic) single-photon subtraction is not efficiently simulatable if the squeezing of the input is sufficiently high. Finally, we tackled the harder computation of the $(s_1,s_2)$-ordered function of the cubic gate to showcase the power of our method to provide quantitative answers also for nontrivial non-Gaussian gates. We emphasize that, in the most general setting, we always assume that definite sets of gates, input states and measurements have been fixed and their $(s)$-ordered QPDs have been characterized. Even when those steps imply a non-negligible numerical cost, it is crucial that this is needed only once for each type of element (state, gate or measurement) and it can thus be considered as a constant overhead to the simulation cost.

\subsection{Loss tolerance of photonic quantum gates and their resourcefulness}
The presented techniques and results can not only directly be applied to existing photonic quantum circuits to test their simulability beyond the requirement of Wigner negativity, but they also probe the robustness of different circuit designs to losses. They allow us to pinpoint the most sensitive gates in the circuit and provide constraints for the quality of technological components that implement them. For example, taking Figs. \ref{fig:squeezing} and \ref{fig:photsubtrparam} for the squeezing and beam-splitter channels, respectively, we can interpret the minimal value of $s_{i}$ attained by each curve (for $s_{i+1}=-1$) as quantifying the maximum amount of losses above which said gate can always be represented by a regular PDF, no matter the input state.

\subsection{Outlook}

Our results provide a significant step forward towards the goal of singling out the salient quantum features needed to hinder any attempt of simulating a photonic quantum process. Many directions remain to be explored, including but not limited to: the role of entanglement at preventing larger combinations of quantum gates, acting on fewer modes each, to be simulated in a single shot; the implications of the conditions on $(s)$-ordered quasiprobability distributions on the existence of local hidden-variable models; the relation between common measures of Wigner negativity and the nonclassical depth of Wigner-negative states, to determine which one best captures the quantitative aspects of this resource.\\

In particular, concerning the role of entanglement, one may try to regroup gates whose transfer functions cannot be made individually positive into larger ones, but if these gates act on more than one mode and can thus create entanglement, this requires sampling from a probability distribution over a larger set of variables, increasing the sampling costs accordingly.

It is also interesting to explore what happens when we drop some of the computational assumptions. A first key step is to connect our approach to the problem of compiling. In the present work, we have assumed that the decomposition of the system into a sequence of fundamental gates is known. In general, there may be many different decompositions that lead to the same global operation, a problem which concerns more the \emph{compiling} aspect of the algorithmic design. Nevertheless, to each of these instances one can readily apply our procedure to determine the resourcefulness of each gate independently of the specific instance itself. 

One may also argue that the computational cost of sampling from a probability distribution is not really taken into account. While from a probability-theory point of view this is more a classical than a quantum problem, it is clear that quantum properties of states, gates, and measurements, may lead to highly oscillating PDFs on phase space. These features may render our simulation protocol extremely costly, even when Algorithm \ref{alg:main} would technically succeed. Moreover, the sampling cost has been explored through the lens of (s)-ordered quasiprobabilities in \cite{lim2023approximating}. Given sufficient classicality, the output parameter can be made arbitrarily close to one for a single step. This suggests that the step in question may be simulatable in polynomial time. Furthermore, allowing for additive errors in the sampling allows for more space in the ordering parameter's optimisation algorithm.

Finally, one may want to challenge altogether the idea that one cannot construct a sampling procedure if the phase space representation manifests negativity. Indeed, it has been shown that we can use sampling techniques to deal with non-positive quasi-probabilities by paying an additional sampling cost \cite{pashayan2015estimating} that is proportional to the negativity volume. Some recent work on the role of phase space negativity on quantum kernel methods \cite{chabaud2024phasespacenegativitycomputationalresource} may help us get a more nuanced idea of the role of the negativity volume in cases where our Algorithm \ref{alg:main} fails.

\begin{acknowledgments}
We would like to thank our fellow theoreticians Carlos Lopetegui and Mathieu Isoard for the numerous informative and supportive discussions. We thank Ulysse Chabaud for valuable comments on the manuscript. We acknowledge financial support from the ANR JCJC project NoRdiC (ANR-21-CE47-0005), the Plan France 2030 through the project OQuLus (ANR-22-PETQ-0013), and the HORIZON-EIC-2022- PATHFINDERCHALLENGES-01 programme under Grant Agreement Number 101114899 (Veriqub).
\end{acknowledgments}

\bibliographystyle{quantum}
\bibliography{biblio}

\onecolumn\newpage
\appendix

\renewcommand*\thefigure{\thesection\arabic{figure}}
\section{Explicit calculations for Eq. (\ref{proba_distrib_sequence})}
\label{app:A}
Similarly to its construction for operators, it is possible to construct ordered quasiprobability distributions for quantum channels (see e.g. \cite{rahimi2016sufficient}). Figure~\ref{fig:inter} illustrates a generic quantum process; the simulation task would amount to sampling from $p(\bm{n}) = \Tr[\Rh_{\mathrm{out}} \Pi_{\bm{n}}]$. It is possible to rewrite this probability distribution in terms of $(\bm{s})$-ordered quasi-probability distributions using a generalization of the trace rules for Wigner functions \cite{cahill1969ordered}: for any two bounded operators $\hat{O}_1$, $\hat{O}_2$ of trace class we can write 
    \begin{equation}
        \Tr[\hat{O}_1 \hat{O}_2] = \int_{\bm{\xi} \in \mathbb{C}} \frac{d^{2M}\bm{\xi}}{\pi^M} W_{\hat{O}_1}^{(\bm{-s})}W_{\hat{O}_2}^{(\bm{s})}.
    \end{equation}
    Thus leading to
    \begin{equation}
        \label{proba_first}
        p(\bm{x}) = \Tr[\Rh_{\mathrm{out}} \Povm_{\bm{x}}] = \pi^{2M} \int_{\bm{\beta} \in \mathbb{C}} d^{2M}\bm{\beta} \ W_{\Povm_{\bm{x}}}^{(-\bm{s})}(\bm{\beta})W_{\Rh_{\mathrm{out}}}^{(\bm{s})}(\bm{\beta}),
    \end{equation}
    where
    \begin{equation}
        W_{\Povm_{\bm{x}}}^{(-\bm{s})}(\bm{\beta}) = \int_{\bm{\xi} \in \mathbb{C}} \frac{d^{2M}\bm{\xi}}{\pi^{2M}} \Tr[\Povm_{\bm{n}}\hat{D}_{-\bm{s}}(\bm{\xi})]e^{\bm{\beta} \bm{\xi}^\dag - \bm{\xi} \bm{\beta}^\dag}
    \end{equation}
    and 
    \begin{equation}
        W_{\Rh_{\mathrm{out}}}^{(\bm{s})}(\bm{\beta}) = \int_{\bm{\xi} \in \mathbb{C}} \ \frac{d^{2M}\bm{\zeta}}{\pi^{2M}}\chi_{\Rh_{\mathrm{out}}}^{(\bm{s})}(\zeta)e^{\bm{\beta}\bm{\zeta}^\dag - \bm{\zeta}\bm{\beta}^\dag}.
    \end{equation}
    One can write the $M$ modes input state in terms of displacement operators as
    \begin{equation}
        \Rh_{\mathrm{in}} = \int_{\bm{xi} \in \mathbb{C}}\chi_{\Rh_{\mathrm{in}}}^{(\bm{s'})}(\bm{\xi}) e^{-\bm{\xi}\bm{s'}\bm{\xi}^\dag/2}\hat{D}^\dag(\bm{\xi})\frac{d^{2M}\bm{\xi}}{\pi^M}.
    \end{equation}
    This implies, by linearity of the quantum process, that the output state is 
    \begin{equation}
        \label{out_char}
        \Rh_{\mathrm{out}} =  \int_{\bm{\xi} \in \mathbb{C}}\chi_{\Rh_{\mathrm{in}}}^{(\bm{s'})}(\bm{\xi}) e^{-\bm{\xi}\bm{s'}\bm{\xi}^\dag/2}\mathcal{E}(\hat{D}^\dag(\bm{\xi}))\frac{d^{2M}\bm{\xi}}{\pi^M}.
    \end{equation} 
    Moreover, by virtue of the inverse Fourier transform, the $(\bm{s})$-ordered characteristic function of the output state is
    \begin{equation}
        \chi_{\Rh_{\mathrm{out}}}^{(\bm{s})}(\bm{\zeta}) = \Tr[\Rh_{\mathrm{out}}\hat{D}_{\bm{s}}(\bm{\zeta})]
    \end{equation}
    By plugging Eq. (\ref{out_char}) into the last we find 
    \begin{equation}
        \chi_{\Rh_{\mathrm{out}}}^{(\bm{s})}(\bm{\zeta})  = \int_{\bm{\xi} \in \mathbb{C}}\chi_{\Rh_{\mathrm{in}}}^{(\bm{s'})}(\bm{\xi}) \Tr[\mathcal{E}(\hat{D}^\dag_{\bm{s'}}(\bm{\xi}))\hat{D}_{\bm{s}}(\bm{\zeta})]\frac{d^{2M}\bm{\xi}}{\pi^M}
    \end{equation}
    Which can be re-written in terms of the $(\bm{s'})$-ordered Wigner function of the input state 
    \begin{equation}
        \chi_{\Rh_{\mathrm{out}}}^{(\bm{s})}(\bm{\zeta}) \ = \ \int_{\bm{\xi} \in \mathbb{C}}\frac{d^{2M}\bm{\xi}}{\pi^{2M}} \int_{\bm{\alpha} \in \mathbb{C}} d^{2M}\bm{\beta} \  W_{\Rh_{\mathrm{in}}}^{(\bm{s'})}(\bm{\beta})  \Tr[\mathcal{E}(\hat{D}^\dag_{\bm{s'}}(\bm{\xi}))\hat{D}_{\bm{s}}(\bm{\zeta})]. 
    \end{equation}
    Using the Fourier transform we can find a relation between the input and output ordered Wigner functions,
    \begin{equation}
        W_{\Rh_{\mathrm{out}}}^{(s)}(\bm{\beta}) = \int_{\bm{\alpha} \in \complex} d^{2M}\bm{\alpha} \ W_{\Rh_{\mathrm{in}}}^{(\bm{s'})}(\bm{\alpha}) \Lambda_{\mathcal{E}}^{(-\bm{s'},\bm{s})}(\bm{\alpha},\bm{\beta}) 
    \end{equation}
    where we define the $(\bm{s'},\bm{s})$-ordered transfer function $\Lambda_{\mathcal{E}}^{(\bm{s'},\bm{s})}(\alpha,\beta)$ as 
    \begin{equation}
    \Lambda_{\mathcal{E}}^{(\bm{s'},\bm{s})}(\bm{\alpha},\bm{\beta}) = \int_{\bm{\zeta} \in \mathbb{C}} \frac{d^{2M}\bm{\zeta}}{\pi^M} \  \int_{\bm{\xi} \in \mathbb{C}} \frac{d^{2M}\bm{\xi}}{\pi^{2M}} \ e^{\bm{\beta} \bm{\zeta}^\dag - \bm{\zeta}\bm{\beta}^\dag+\bm{\xi}\bm{\alpha}^\dag - \bm{\alpha} \bm{\xi}^\dag}   e^{\bm{\zeta}\bm{s}\bm{\zeta}^\dag/2 + \bm{\xi}\bm{s'}\bm{\xi}^\dag/2} \Tr[\mathcal{E}(\hat{D}^\dag(\bm{\xi}))\hat{D}(\bm{\zeta})] .
    \end{equation}
    Finally, by plugging last equation into Eq. (\ref{proba_first}) we find the highlighted result of \cite{rahimi2016sufficient}:
    \begin{equation}
        p(\bm{x}) = \iint_{\bm{\alpha}, \bm{\beta} \in \mathbb{C}} \pi^{M}d^{2M}\bm{\alpha}d^{2M}\bm{\beta}\  W_{\Rh_{\mathrm{in}}}^{(\bm{s'})}(\bm{\alpha}) \Lambda_{\mathcal{E}}^{(-\bm{s'},\bm{s})}(\bm{\alpha},\bm{\beta}) W_{\Povm_{\bm{x}}}^{(-\bm{s})}(\bm{\beta})
    \end{equation}
    The last step needed to find Eq. (\ref{proba_distrib_sequence}) is to replace $\mathcal{E}$ by a sequence of trace-preserving quantum channels $\mathcal{E}_1 \circ \mathcal{E}_2 \circ \dots \circ \mathcal{E}_k$ and notice that 
    \begin{equation}
        \Lambda_{\mathcal{E}_1 \circ \mathcal{E}_2 }^{(-\bm{s'},\bm{s})}(\bm{\alpha},\bm{\beta}) = \int_{\bm{\gamma} \in \mathbb{C}} \Lambda_{\mathcal{E}_1}^{(-\bm{s'},\bm{s''})}(\bm{\alpha},\bm{\gamma})\Lambda_{\mathcal{E}_2 }^{(-\bm{s''},\bm{s})}(\bm{\gamma},\bm{\beta})
    \end{equation}
    Leading naturally to Eq. (\ref{proba_distrib_sequence}).

These functions are all defined on the complex plane, rather than on the real symplectic phase-space that is commonly encountered when dealing with Wigner functions. To convert to quadrature notation, one can use the following identities:
\begin{equation}
   \left\{  \begin{aligned}
        & \hat{\bm{q}} = \hat{\bm{a}} + \hat{\bm{a}}^\dagger \ , \ \ \ \ \hat{\bm{p}} = -i (\hat{\bm{a}} - \hat{\bm{a}}^\dagger) \\
        & \vec{q} = \dfrac{\bm{\xi} + \bm{\xi}^\dagger}{2} \ , \ \ \ \ \vec{p} = \dfrac{\bm{\xi} - \bm{\xi}^\dagger}{2 i }
    \end{aligned}  \right.
\end{equation}

\section{Explicit calculations for the cubic gate}
\label{app:B}

Let $\hat{U}_\gamma = \exp \left(- i \frac{\gamma}{3!} \hat{q}^3 \right)$. Firstly, we need to compute the characteristic function of this gate:
\begin{equation}
    \Tr \left[  \hat{U}_\gamma \hat{D}^\dagger ( \xi ) \hat{U}^{\dagger}_\gamma  \hat{D} (\zeta ) \right]
\end{equation}
It is convenient to sart by rewriting $D(\xi)$ in terms of the real, quadratures variables:
\begin{equation}
\hat{D}(\xi) \ = \ \exp \left\{ \xi \hat{a}^\dagger - \xi^{*} \hat{a}  \right\} \ = \ \exp \left\{ -i ( q_{1} \hat{p} - p_{1}  \hat{q}  )  \right\}
\end{equation}
We have that $\hat{U}_\gamma  \hat{q}  \hat{U}^{\dagger}_\gamma   = \hat{q}$, while $\hat{U}_\gamma \hat{p}  \hat{U}^{\dagger}_\gamma  = \hat{p}  + \gamma \hat{q}^2$, from which it follows that:
\begin{equation}
    \hat{U}_\gamma \hat{D}^\dagger ( \xi ) \hat{U}^{\dagger}_\gamma \ = \ \exp \left\{ i \left[ q_{1} ( \hat{p} + \gamma \hat{q}^2 ) - p_{1}  \hat{q}  \right] \right\}
\end{equation}
where $q_{1} = \Re \xi$ and $p_{1} = \Im \xi$. 
To compute the trace using the eigenbasis of $\hat{q}$, it is useful to rearrange this result as: 
\begin{equation}
\label{eq:cubicgate2}
     \exp \left\{ i \left[ q_{1} ( \hat{p} + \gamma \hat{q}^2 ) - p_{1}  \hat{q}  \right] \right\} \ = \ e^{a \hat{p} } e^{b \hat{q} + c \hat{q}^2 + d }
\end{equation}

for a proper choice of $a,b,c \in \complex$. Notice that there must be a solution, since commutators of quadratic operators in the quadratures form a closed subalgebra and a term in $\hat{p}^2$ cannot appear, as it is not present in the exponent on the left-hand side. We can compute $a,b,c$ by applying the Baker-Campbell-Hausdorff formula to the right-hand side, to get:
\begin{equation}
\begin{aligned}
    & e^{a \hat{p} } e^{b \hat{q} + c \hat{q}^2} \  \\
    & = \ \exp \left\{ a \hat{p} + b \hat{q} + c \hat{q}^2 + \frac{1}{2} [ a \hat{p} , b \hat{q} + c \hat{q}^2  ]  + \dots \right\} \  \\ 
    & = \ \exp \left\{ a \hat{p} + (b - 2 i ac) \hat{q} + c \hat{q}^2 - i a b   + \frac{1}{12} [ a \hat{p} , - 4 i a c \hat{q} ]  \right\} \  \\
    & = \ \exp \left\{ a \hat{p} + (b -2 i  ac) \hat{q} + c \hat{q}^2 - i a b  -  \frac{2}{3} a^2 c  \right\}
\end{aligned}
\end{equation}
and we used $[ \hat{q}, \hat{p} ] = 2 i $. All higher order commutators vanish, since the last second-order nested commutator is a scalar. By direct comparison with the left-hand side of Eq. (\ref{eq:cubicgate2}) we conclude that:
\begin{equation}
    \begin{aligned}
        & a =  i q_{1}  \ , \ \ c =  i  q_{1} \gamma \\ 
        & b = -i  p_{1} - 2 i q_{1}^{2} \gamma \ , \ \ d = i q_{1} p_{1}  + \frac{4}{3} i  \gamma  q_{1}^{3} 
    \end{aligned}
\end{equation}
If we similarly rewrite $\hat{D}(\zeta) = e^{ i q_{2} p_{2} } e^{-i q_{2} \hat{p} } e^{ i p_{2} \hat{q} }$, with $q_{2} = \Re \zeta$ and $p_{2} = \Im \zeta$, we finally have:
\begin{equation}
\begin{aligned}
    & \Tr \left[  \hat{U}_\gamma \hat{D}^\dagger ( \xi ) \hat{U}^{\dagger}_\gamma  \hat{D} (\zeta ) \right] \ = \\
    & = \ e^{ \phi } \int_{q \in \mathbb{R}} dq \ \langle q \vert e^{a \hat{p} } e^{b \hat{q} + c \hat{q}^2 } e^{-i q_{2} \hat{p} } e^{ i p_{2} \hat{q} } \vert q \rangle \ = \\
    & = \ e^{ \phi } \int_{q \in \mathbb{R}} dq \ e^{ i p_{2} q } \langle q + 2  q_{1} \vert e^{b \hat{q} + c \hat{q}^2 }  \vert q + 2 q_{2} \rangle \  = \\ 
& = \ \frac{1}{2} e^{ \phi }  \delta (q_{1} - q_{2} ) \int_{q \in \mathbb{R}} dq \  e^{ i p_{2} q + b(q + 2 q_{1}) + c(q+ 2 q_{1})^2}  \ = \\
& = \ \frac{1}{2} e^{ \tilde{\phi} }  \delta (q_{1} - q_{2} ) \int_{q \in \mathbb{R}} dq \  e^{i (p_2 - p_{1}  - 2  \gamma q_{1}^{2} ) q  + i \gamma  q_1 q^2}
\end{aligned}
\end{equation}
with $\phi =i (q_{1} p_{1} + q_{2} p_{2} )  + \frac{4}{3} i \gamma q_{1}^3$ and $\tilde{\phi} = \textcolor{red}{-}i  q_{1} ( p_{1} - p_{2} )  + \frac{4}{3} i \gamma q_{1}^3$. The remaining integration is the Fourier transform of a Gaussian with an imaginary variance, and it can computed by regularizing the integral, to find:
\begin{equation}
     \int_{q \in \mathbb{R}} dq \  e^{i (p_2 - p_{1}  -2  \gamma q_{1}^{2} ) q  + i \gamma  q_1 q^2}  = \sqrt{\dfrac{ i \pi}{ \gamma q_{1} } } e^{-i \frac{(p_2 - p_{1} -2  \gamma q_{1}^{2} )^2 }{4 \gamma q_{1}} }
\end{equation}

Putting everything together, we arrive at:
\begin{equation}
\label{eq:charfunctcubic}
        \Tr \left[  \hat{U}_\gamma \hat{D}^\dagger ( \xi ) \hat{U}^{\dagger}_\gamma \hat{D} (\zeta ) \right]  \ = \ \frac{1}{2} \delta (q_{1} - q_{2} )  \sqrt{\dfrac{  i \pi}{ \gamma q_{1} } } e^{- \frac{i }{4 \gamma q_{1}}(p_1 - p_{2})^2  + \textcolor{red}{\frac{1}{3} } i \gamma q_{1}^3 }
\end{equation}
The next steps consist in performing the Fourier transform of this function with respect to $q_1, q_2, p_1, p_2$, with additional Gaussian weights that impose the $(s_{1},s_{2})$ ordering, that is, multiplying the result of Eq. (\ref{eq:charfunctcubic}) by:
\begin{equation}
    e^{\alpha^* \xi - \alpha \xi^* + \beta \zeta^* - \beta^* \zeta} e^{\frac{ s_{1} \vert \xi \vert^{2}}{2} + \frac{ s_{2} \vert \zeta \vert^{2}}{2}  } \  = \ \exp \left[ 2 i ( q_\alpha p_{1} - p_\alpha q_1 + p_\beta q_2 - q_\beta p_2 )  \right] \exp \left[\frac{s_1}{2} ( q_{1}^{2} + p_{1}^{2})+ \frac{s_2}{2} ( q_{2}^{2} + p_{2}^{2}) \right]
\end{equation}

Because of the delta distribution, the integration with respect to $q_{2}$ is trivial. Also, the integrations with respect to $p_{1}$ and $p_{2}$ amounts to computing the Fourier transform of Gaussian functions with complex variance; these converge if and only if the real part of the variance is positive, such that the integrand decays exponentially in $p_{1}$ and $p_{2}$. Also, observing that the Gaussian factor in $\Tr \left[  \hat{U}_\gamma \hat{D}^\dagger ( \xi ) \hat{U}^{\dagger}_\gamma  \hat{D} (\zeta ) \right]$ is purely imaginary, the conditions of convergence are just $s_{1} < 0$ and $s_{2} < 0$. The general formula that should be applied for the integrations in $p_{1}$ and $p_{2}$ is then:

\begin{equation}
   \forall A, B \in \complex: \  \Re ( A ) > 0 \  \ \ \implies \ \ \int_{x \in \mathbb{R}} e^{ - A x^2 + i B x} dx \ = \ \sqrt{\dfrac{\pi}{A}} e^{ - \dfrac{B^2}{4A}}
\end{equation}
and the end result is:
\begin{multline}
   \Lambda^{(s_1, s_2 )}_{\hat{U}_\gamma} (q_\alpha, p_\alpha, q_\beta, p_\beta)  \ \ = \ \  \int_{q_{1} \in \mathbb{R}} \dfrac{d q_{1}}{\sqrt{2\pi}} \   \sqrt{\dfrac{ i}{2 \gamma q_{1} s_{1} s_{2}- i (s_1+s_2) } } \\
    \times \exp \left[   \dfrac{  4 \gamma q_1 (s_{2} q_\alpha^2 + s_{1} q_\beta^2 )- 2i (q_\alpha - q_\beta)^2 + }{ 2 \gamma q_{1} s_{1} s_{2}- i (s_1 + s_2)  } + \frac{s_{1}+s_{2}}{2} q_{1}^2 + \frac{i}{3}  \gamma q_{1}^{3} + 2 i (p_{\beta} - p_{\alpha} ) q_{1}   \right]
\end{multline}
Again, for $s_{1} < 0$ and $s_2 < 0$ the integral must converge. Notice, moreover, that the integrand is Gaussian in $q_{\alpha}, q_{\beta}$ and the real part of the variance is positive as long as $s_{1}, s_{2} < 0$. Finally, it depends only upon the combination $p_{\alpha} - p_{\beta}$ and not on the single variables independently. We can then conclude that $\Lambda^{(s_{1}, s_{2} )}_{\hat{U}_\gamma} (q_\alpha, p_\alpha, q_\beta, p_\beta)$ is a well-defined function whenever $s_{1}, s_{2} < 0$ and, by Theorem \ref{thm:nGunitaries}, we know that it cannot be positive-semidefinite for any $s_{1}> -1$ and $s_{2} > -1$. At the same time, it must be positive semidefinite for $s_{1} = s_{2} = -1$, since it then reduces to:
\begin{equation}
     \Lambda^{(-1, -1 )}_{\hat{U}_\gamma} (q_\alpha, p_\alpha, q_\beta, p_\beta) \ \propto \ \vert \langle \beta \vert \hat{U}_\gamma \vert \alpha \rangle \vert^2 
\end{equation}
A somewhat simpler expression for $ \Lambda^{(s_1, s_2 )}_{\hat{U}_\gamma}$ for $s_1,s_2 < 0$ can be computed also using Eq. (\ref{eq:Conv}) for the change of the $s$-parameters starting with  $\Lambda^{(0, 0 )}_{\hat{U}_\gamma}$, which can be directly computed to be:
\begin{equation}
    \Lambda^{(0, 0 )}_{\hat{U}_\gamma} ( q_\alpha , p_\alpha , q_\beta , p_\beta) \ = \ \frac{2}{\pi} \vert\gamma\vert^{-1/3}\delta(q_\alpha-q_\beta) \mathrm{Ai} \left( \dfrac{2(p_\beta - p_\alpha)}{\gamma^{1/3}}  + 4 \vert\gamma\vert^{2/3} q_{\alpha}^{2} \right)
\end{equation}
one can then finally get the following form for $s_1, s_2 < 0$:

\begin{multline}
     \Lambda^{(s_1, s_2)}_{\hat{U}_\gamma} ( q_\alpha , p_\alpha , q_\beta , p_\beta) \ = \\ 
     = 4 \iint \frac{d{q'}_\alpha d{p'}_\alpha d {q'}\beta d{p'}_\beta }{\pi^2 s_1 s_2 } e^{2\frac{(q_\alpha - {q'}_{\alpha})^2+ (p_\alpha - {p'}_{\alpha})^2}{s_1} + 2\frac{(q_\beta - {q'}_{\beta})^2+ (p_\beta - {p'}_{\beta})^2}{s_2} }\Lambda^{(0, 0 )}_{\hat{U}_\gamma} ( {q'}_\alpha , {p'}_\alpha , {q'}_\beta , {p'}_\beta) \ = \\
     = \ \frac{4}{\pi^{5/2} \vert \gamma\vert^{1/3} \Sigma } e^{-\frac{2(q_\alpha - {q}_{\beta})^2}{\Sigma}} \int dx \sqrt{\frac{2}{\sigma}} e^{-\frac{2}{\sigma} \left( x - \mu  \right)^2} \int d u \ e^{-\frac{2(u-p)^2}{\Sigma}}  \mathrm{Ai} \left( \frac{2 u}{\gamma^{1/3}}+ 4 \vert \gamma \vert^{2/3} x^2 \right) 
\end{multline}
where $p= p_\alpha - p_\beta$, $\Sigma = \vert s_1 + s_2 \vert$, $\sigma = \frac{s_1 s_2}{\vert s_1 + s_2\vert}$ and $\mu = \frac{s_2 q_\alpha + s_{1} q_\beta}{\Sigma}$. Notice that this is a double Gaussian smoothing of the Airy function. Due to the highly oscillatory behavior of the latter, however, it is still very challenging to find conditions on $(s_1,s_2)$ as functions of $\gamma$ such that the final result is everywhere non-negative.

\end{document}